\documentclass[onecolumn]{aa}            
\usepackage{txfonts}                    

\usepackage{hyperref}        
\usepackage{natbib}
\usepackage{amsmath,amssymb}

\usepackage{graphicx}
\usepackage{ulem}
\usepackage{bm}            
\usepackage{mathtools}

\usepackage{xspace}

\usepackage{makecell}
\usepackage{multirow}

\usepackage{xcolor}

\newcommand{\revise}[1]{{#1}} 

\newcommand{\xM}{\ensuremath{x_{\rm M}}}
\newcommand{\yM}{\ensuremath{y_{\rm M}}}
\newcommand{\Pfit}{\ensuremath{P_{\rm fit}}}
\newcommand{\taufit}{\ensuremath{\tau_{\rm fit}}}
\newcommand{\Omgfit}{\ensuremath{\Omega_{\rm fit}}}
\newcommand{\gamfit}{\ensuremath{\gamma_{\rm fit}}}
\newcommand{\omgA }{\ensuremath{\omega_{\rm A}}}
\newcommand{\omgAe}{\ensuremath{\omega_{\rm Ae}}}
\newcommand{\omgAi}{\ensuremath{\omega_{\rm Ai}}}
\newcommand{\vA }{\ensuremath{v_{\rm A}}}
\newcommand{\vAi}{\ensuremath{v_{\rm Ai}}}
\newcommand{\vAe}{\ensuremath{v_{\rm Ae}}}
\newcommand{\Rtrd}{\ensuremath{\mathcal{R}}}
\newcommand{\ltrd}{\ensuremath{\ell}}
\newcommand{\ck}{\ensuremath{c_{\rm kink}}}
\newcommand{\Lph}{\ensuremath{L_{\rm ph}}}
\newcommand{\Taxial}{\ensuremath{T_{\rm axial}}}
\newcommand{\phiac}{\ensuremath{\phi_{\rm ac}}}
\newcommand{\Nextrm}{\ensuremath{N_{\rm extrm}}}
\newcommand{\Pobs}{\ensuremath{P_{\rm obs}}}
\newcommand{\tauPobs}{\ensuremath{(\tau/P)_{\rm obs}}}

\defcitealias{1983SoPh...88..179E}{ER83}
\defcitealias{2002ApJ...577..475R}{RR02}
\defcitealias{2008ApJ...676..717L}{L08}
\defcitealias{2008ApJ...679.1611T}{T08}
\defcitealias{2015A&A...582A.120S}{SL15}

\newcommand{\uvec}[1]{\vec{e}_{#1}}
\newcommand{\Exp}[1]{\ensuremath{{\rm e}^{#1}}}

\newcommand{\Alf}{Alfv$\acute{\rm e}$n}
\newcommand{\Alfvenic}{Alfv$\acute{\rm e}$nic}

\newcommand{\mathd}{\ensuremath{{\rm d}}}

\newcommand{\rhoi}{\ensuremath{\rho_{\rm i}}}
\newcommand{\rhoe}{\ensuremath{\rho_{\rm e}}}

\begin{document}

\title{Damped kink motions in a system of two solar coronal tubes with elliptic cross-sections}
\author{      Mijie Shi     \inst{1}
    	\and  Bo Li         \inst{1}
        \and  Shaoxia Chen  \inst{1}
        \and  Hui Yu        \inst{1}
        \and  Mingzhe Guo   \inst{1,2} 
       }

\institute{
Shandong Provincial Key Laboratory of Optical Astronomy and 
    Solar-Terrestrial Environment,
Institute of Space Sciences, Shandong University, Weihai 264209, China\\
    \email{bbl@sdu.edu.cn}
    \and 
Centre for Mathematical Plasma Astrophysics, 
Department of Mathematics, KU Leuven, 
Celestijnenlaan 200B, B-3001 Leuven, Belgium
}

\titlerunning{Damped kink motions in two-elliptic-tube systems}
\authorrunning{Shi et al.}

\date{Received ......... / Accepted .........}

\abstract
{}
{This study is motivated by observations of coordinated transverse displacements in
   neighboring solar active region loops, addressing specifically how the behavior of kink motions in straight two-tube equilibria is impacted by tube interactions and tube cross-sectional shapes.}
{We work with linear, ideal, pressureless magnetohydrodynamics. 
Axially standing kink motions are examined as an initial value
   problem for transversely structured equilibria involving two identical, field-aligned, density-enhanced tubes with elliptic cross-sections (elliptic tubes). 
Continuously nonuniform layers are implemented around both tube boundaries. 
We numerically follow the system response to external velocity drivers, 
   largely focusing on the quasi-mode stage of internal flows to derive the pertinent periods and damping times.  
}
{The periods and damping times we derive for two-circular-tube setups
   justify available modal results found with the T-matrix approach. 
Regardless of cross-sectional shapes, our nonuniform layers feature the
   development of small-scale shears and energy accumulation around \Alf\ resonances, indicative of resonant absorption and phase-mixing. 
As with two-circular-tube systems, our configurational symmetries
   make it still possible to classify lower-order kink motions by the
   polarization and symmetric properties of the internal flows; hence such mode labels as $S_x$ and $A_x$.   
However, the periods and damping times for two-elliptic-tube setups
   further depend on cross-sectional aspect ratios, with $A_x$ motions occasionally damped less rapidly than $S_x$ motions. 
We find uncertainties up to $\sim 20\%$ ($\sim 50\%$)
   for the axial \Alf\ time (the inhomogeneity lengthscale)
   if the periods (damping times) computed for two-elliptic-tube setups are 
   seismologically inverted with canonical theories for isolated circular tubes.
}
{The effects of loop interactions and cross-sectional shapes need to be considered
   when the periods and particularly the damping times are seismologically exploited for coordinated transverse displacements in adjacent coronal loops. 
}


\maketitle 

\section{Introduction}
\label{sec_intro}
Cyclic transverse displacements of solar coronal loops are arguably 
   the most extensively observed collective motion in modern solar coronal seismology \citep[see e.g.,][for recent reviews]{2020ARA&A..58..441N,2021SSRv..217...73N}.
Two regimes have been established.
The decayless regime is such that the displacements
   show little damping and their magnitudes are usually substantially smaller than visible loop widths, with both features already clear when this regime was first identified in measurements with Hinode \citep{2012ApJ...759..144T} and the Solar Dynamics Observatory/Atmospheric Imaging Assembly \citep[SDO/AIA;][]{2012ApJ...751L..27W,2013A&A...552A..57N,2013A&A...560A.107A}.
This regime is known not to be connected with
   eruptive events but ubiquitous in active region (AR) loops, 
   as evidenced by a statistical survey of the SDO/AIA data \citep{2015A&A...583A.136A} and by the recent series of analyses of the measurements with the Extreme Ultraviolet Imager (EUI) on board the Solar Orbiter \citep[e.g.,][]{2022MNRAS.516.5989Z,2023ApJ...946...36P,2023NatCo..14.5298Z}.
Decaying loop displacements, on the other hand, 
   typically damp over several cycles and are of larger amplitudes as revealed by their first imaging observations by the Transition Region and Coronal Explorer \citep[TRACE;][]{1999SoPh..187..261S,1999ApJ...520..880A,1999Sci...285..862N}.
These decaying motions are usually associated with
   lower coronal eruptions \citep{2015A&A...577A...4Z}.
Regardless, there exist ample detections of decaying displacements,
   an inexhaustive list being those by Hinode \citep[][]{2008A&A...482L...9O,2008A&A...487L..17V,2008A&A...489L..49E}, the Solar TErrestrial RElations Observatories \citep[STEREO;][]{2009ApJ...698..397V},
   and by SDO/AIA \citep{2011ApJ...736..102A,2012A&A...537A..49W}.
Statistical studies therefore prove possible either through compiling published 
   results \citep[e.g.,][]{2013A&A...552A.138V} or via directly cataloging the SDO/AIA events \citep{2016A&A...585A.137G,2019ApJS..241...31N}.
Decaying cyclic displacements were specifically established to be
   nearly exclusively axial fundamentals 
   (e.g., Fig.9 in \citealt{2016A&A...585A.137G} and Fig.5 in \citealt{2019ApJS..241...31N}).

Seismological applications of cyclic transverse displacements typically start
   with their identification as trapped fast kink motions.
Practically, this identification largely relies on the scheme for
   classifying collective motions in straight, static, field-aligned configurations 
   where isolated density-enhanced tubes with circular cross-sections (circular tubes hereafter) are embedded in an otherwise uniform corona 
   (\citealt{1983SoPh...88..179E}, \citetalias{1983SoPh...88..179E}; 
   also \citealt{1975IGAFS..37....3Z,1986SoPh..103..277C}).
That the configuration is structured only transversely and
   in a one-dimensional (1D) manner means that the azimuthal wavenumber $m$ makes physical sense, with kink motions corresponding to $m=1$ 
   (see the textbooks by \citealt{2019CUP_Roberts} and \citealt{2019CUP_goedbloed_keppens_poedts}).
Let ``\citetalias{1983SoPh...88..179E} equilibria'' label specifically those where the
    transverse structuring is piecewise constant.
The relevant theories then enable the measured periods
    of axially standing kink motions to be employed to infer the axial \Alf\ time and hence the coronal magnetic field strength for decayless \citep[e.g.,][]{2019ApJ...884L..40A,2023ApJ...944....8L} and decaying regimes alike 
    (\citealt{2001A&A...372L..53N}; also the reviews by e.g., \citealt{2005LRSP....2....3N,2012RSPTA.370.3193D,2021SSRv..217...73N}).
While undamped for \citetalias{1983SoPh...88..179E} equilibria, 
    trapped fast kink motions are in general
    resonantly absorbed in the \Alf\ continuum when the 1D structuring is allowed to be continuous \citep[][and references therein]{2002ApJ...577..475R,2002A&A...394L..39G}. 
Theoretically, the concept of kink quasi-modes then arises 
    and the internal kink motions are damped in conjunction with the accumulation and phase-mixing of localized \Alfvenic\ motions (e.g., \citealt{1991PhRvL..66.2871P,1996ApJ...471..501T,2013ApJ...777..158S}; also the review by \citealt{2011SSRv..158..289G}).
Seismologically, resonant absorption has been customarily invoked to interpret
    the decay of large-amplitude loop displacements, thereby allowing the measured damping times to be employed to deduce such key parameters as the transverse inhomogeneity lengthscales \citep[e.g.,][]{2003ApJ...598.1375A,2008A&A...484..851G,2011ApJ...740...44A,2015ApJ...811..104A,2022FrASS...926947A}.

Deviations from the canonical \citetalias{1983SoPh...88..179E} equilibria are
    known to impact the behavior of collective motions, and we choose to focus on two geometrical properties that render the meaning of the azimuthal wavenumber not as clear \citep[e.g., the review by][]{2020SSRv..216..136L}.   
One concerns loop cross-sections, which may actually be tied to coronal 
    heating via their key role in determining the morphology of coronal loops in, say, soft X-ray \citep{1992PASJ...44L.181K} and EUV \citep{2000SoPh..193...77W} in the first place.
Recent imaging observations with Hi-C \citep{2020ApJ...900..167K} and Hi-C 2.1 \citep{2021ApJ...919...47W}
    suggest that AR loops may maintain a circular cross-section throughout their visible segments.
However, there also exist suggestions that favor elliptic cross-sections to better account for
    the morphology of AR loops, as deduced with the aid of coronal magnetic field modeling \citep[e.g.,][]{1998PASJ...50..111W,2013ApJ...775..120M} and/or    
    multi-vantage-point measurements with STEREO \citep[e.g.,][]{2021ApJ...913...56M}. 
In particular, the derived aspect ratios may readily attain $\sim 1.5-5$ \citep{2013ApJ...775..120M}, 
    a range also compatible with the spectroscopic measurements with Hinode/EUV Imaging Spectrometer \citep[EIS;][]{2019ApJ...885....7K}. 
Restrict ourselves to straight configurations where coronal loops preserve a constant
     elliptic cross-section along their axes (elliptic tubes hereafter).
Substantial attention has been paid to collective perturbations in such equilibria from both
    the modal \citep[e.g.,][]{2003A&A...409..287R,2009A&A...494..295E,2011A&A...527A..53M,2021ApJ...912...50A} 
    and initial-value-problem (IVP) perspectives \citep{2020ApJ...904..116G}.
Despite the lack of axisymmetry, kink motions remain identifiable 
    as those that transversely
    displace the tube axes, and in general they remain subjected to resonant absorption for continuous transverse structuring.
However, as was first shown by \citet{2003A&A...409..287R}, one now needs to
    discriminate two distinct polarizations, where the internal flows are primarily directed 
    along the major (``major-polarized'' for brevity) and
    minor axes (``minor-polarized''), respectively.
In addition, the periods of major-(minor-) polarized kink motions
    increase (decrease) with the major-to-minor-axis ratio, while the damping times for both polarizations tend to increase with this ratio. 
It was further demonstrated that the density contrast between elliptic tubes and their surrounding fluids
    plays a subtle role in mediating the differences between the dispersive properties of the differently polarized motions \citep{2020ApJ...904..116G}.     
    
Another geometrical property pertains to systems involving multiple tubular structures. 
Observationally, it has long been known that perturbations in such systems may evolve collectively, 
    an incomplete list of examples being the coordinated transverse displacements detected 
    either in groups of prominence threads 
    (e.g., \citealt{1991SoPh..132...63Y,2003SoPh..216..109L,2007Sci...318.1577O}; also the review by 
    \citealt{2018LRSP...15....3A})
    or in neighboring AR loops \citep[e.g.,][]{2000ApJ...537L..69S,2002SoPh..206...69S,2004SoPh..223...77V,2012ApJ...751L..27W,2013ApJ...774..104W}.
Kink motions in straight multi-circular-tube systems have also been extensively examined theoretically.
We choose to concentrate on the modal analyses where a harmonic time-dependence is assumed a priori, 
    noting that multi-dimensional IVP studies prove equally informative 
    \citep[e.g.,][]{2008ApJ...679.1611T,2009ApJ...694..502O,2016ApJ...823...82M,2019ApJ...883...20G}.
The first modal examination was due to
    \citet[][hereafter \citetalias{2008ApJ...676..717L}]{2008ApJ...676..717L},
    who numerically solved the pertinent eigenvalue problem (EVP) for undamped kink modes in 
    a two-tube system. 
Broadly speaking, two different approaches have been employed in further modal studies.
The ensuing EVPs turn out to be analytically tractable in the thin-tube (TT) limit for two-tube equilibria
    when formulated in bi-cylindrical coordinates, for both undamped \citep{2008A&A...485..849V,2010A&A...515A..33R,2023MNRAS.523.2074R}
    and damped kink motions \citep{2011A&A...525A...4R,2014A&A...562A..38G}. 
Similar EVPs have also been formulated with the T-matrix formalism of scattering theory, originally introduced
    to solar contexts by \citet{1985ApJ...298..867B,1987ApJ...312..444B}.
This formalism is sufficiently general for examining kink motions in composite equilibria comprising
    an arbitrary number of circular tubes, with specific investigations available for undamped motions
    in systems with two or three \citep{2009ApJ...692.1582L} up to tens of tubes \citep{2010ApJ...716.1371L,2019A&A...629A..20L}. 
Addressing continuous transverse structuring, 
    the T-matrix approach proves capable of handling the resonant absorption of
    collective motions in general \citep{1994ApJ...436..372K} and particularly that of kink motions
    in two-tube equilibria \citep[][\citetalias{2015A&A...582A.120S} hereafter]{2015A&A...582A.120S}.
The damping is nonetheless assumed to be weak by construction, given that it was incorporated
    in the T-matrix framework exclusively via the thin-boundary (TB) connection formulae
    (see \citetalias{2015A&A...582A.120S} for more details; 
    also see \citealt{1991SoPh..133..227S} for the first derivation of the TB formulae). 
Regardless, kink motions are known to be much more complicated in multi-tube equilibria 
    than for isolated circular tubes, and we restrict ourselves to two-identical-tube systems.
It was shown by \citetalias{2008ApJ...676..717L} (see Fig.2 therein) that 
    kink motions, namely those that displace both tube axes, need to be classified 
    according to both the polarization and symmetric properties of the two internal flows. 
The notions of $S_x$, $A_x$, $S_y$, and $A_y$ then ensue, one subtlety being that 
    the $S_x-A_y$ and $A_x-S_y$ pairs become indistinguishable in terms of frequencies \citep{2008A&A...485..849V} and damping rates (\citetalias{2015A&A...582A.120S}) when the tubes are sufficiently thin.
This subtlety notwithstanding, $A_x$ motions always turn out to oscillate and damp more rapidly 
    than $S_x$ motions: the differences between the frequencies and particularly the damping rates may be   
    substantial enough to impact seismological deductions (see Fig.4 in \citetalias{2015A&A...582A.120S}).

This study is intended to examine damped lower-order kink motions in a two-identical-elliptic-tube system, 
    thereby addressing the situation where neither the equilibrium configuration as a whole nor an individual tube allows the azimuthal wavenumber to make exact sense.
We will adopt linear, pressureless, ideal magnetohydrodynamics (MHD) throughout, given  
    that quiescent AR loops are of primary interest. 
We will adopt an IVP perspective, paying particular attention
    to the frequencies and damping rates of axially standing motions 
    by largely focusing on the duration where the concept 
    of quasi-modes applies.
Our study is new in the following two aspects. 
Firstly, two-elliptic-tube systems have yet to be explored in the context of collective waves, 
    meaning that our study will shed new light on how the dispersive properties of kink motions 
    are impacted by the joint effects of tube interactions and tube cross-sectional shapes. 
When put to seismological contexts, our results can therefore help assess the uncertainties in
    the key physical parameters that one deduces with the customary practice 
    where the joints effects are absent. 
Secondly, there exist no IVP studies dedicated to the dispersive properties of damped kink motions 
    in two-tube systems to our knowledge. 
Our numerical analysis, conducted with a self-developed code, 
    will therefore help verify the T-matrix results obtained by \citetalias{2015A&A...582A.120S}. 

The outline of this manuscript is as follows. 
Section~\ref{sec_prob} offers the mathematical formulation of our IVP together with a description
    of our numerical code. 
We then focus on circular tubes in Sect.\ref{sec_verif}, testing our code outputs against available results
    obtained with independent approaches. 
Section \ref{sec_2ellip} proceeds to present our numerical results for two-elliptic-tube systems.
\revise{Section~\ref{sec_disc} discusses some seismological implications of this study.} 
\revise{Our findings} are summarized in Sect.~\ref{sec_conc}, where some concluding remarks
    are also given.

\section{Problem Formulation and Solution Method}
\label{sec_prob}
\subsection{Governing Equations}
\label{sec_sub_govEq}
We adopt pressureless, ideal MHD as our theoretical framework, 
   in which the primitive dependents
   are the mass density $\rho$, velocity $\vec{v}$, and magnetic field $\vec{B}$. 
Let $(x,y,z)$ be a Cartesian system, and let the equilibrium quantities
   be denoted with a subscript $0$.
We take the equilibrium magnetic field to be uniform and $z$-directed
   ($\vec{B}_0 = B_0 \uvec{z}$).
Only straight, static, field-aligned configurations are of interest, 
   meaning that $\vec{v}_0 = 0$ 
   and the structuring is encapsulated in $\rho_0(x,y)$.
The \Alf\ speed $\vA$ is defined by
   $\vA^2 = B_0^2/\mu_0\rho_0$,  
   where $\mu_0$ is the magnetic permeability of free space. 
With magnetically closed structures in mind,    
   we place two dense photospheres at $z=0$ and $z=L$.

We formulate an IVP to examine how our equilibrium responds to 
   small-amplitude perturbations (denoted by subscript $1$). 
It follows from linearized, ideal, pressureless MHD equations that    
\begin{eqnarray}
\rho_0\dfrac{\partial v_{1x}}{\partial t}
&=& \dfrac{B_0}{\mu_0}
    \left(\dfrac{\partial B_{1x}}{\partial z}
         -\dfrac{\partial B_{1z}}{\partial x}
    \right), 
            \label{eq_linMHD_3Dvx} \\
\rho_0\dfrac{\partial v_{1y}}{\partial t}
&=& \dfrac{B_0}{\mu_0}
    \left(\dfrac{\partial B_{1y}}{\partial z}
         -\dfrac{\partial B_{1z}}{\partial y}
    \right), 
             \label{eq_linMHD_3Dvy} \\
\dfrac{\partial B_{1x}}{\partial t}
&=& B_0\dfrac{\partial v_{1x}}{\partial z}, 
              \label{eq_linMHD_3DBx} \\
\dfrac{\partial B_{1y}}{\partial t}
&=& B_0\dfrac{\partial v_{1y}}{\partial z}, 
               \label{eq_linMHD_3DBy} \\
\dfrac{\partial B_{1z}}{\partial t}
&=& -B_0
    \left(\dfrac{\partial v_{1x}}{\partial x}
         +\dfrac{\partial v_{1y}}{\partial y}
    \right). 
                    \label{eq_linMHD_3DBz}
\end{eqnarray}
We focus on axially standing motions by adopting the ansatz
\begin{equation}
\label{eq_FourierAnsatz}
\begin{split}
v_{1x}(x,y,z;t) &= \hat{v}_x(x,y;t)\sin(kz), \\
v_{1y}(x,y,z;t) &= \hat{v}_y(x,y;t)\sin(kz), \\    
B_{1x}(x,y,z;t) &= \hat{B}_x(x,y;t)\cos(kz), \\
B_{1y}(x,y,z;t) &= \hat{B}_y(x,y;t)\cos(kz), \\
B_{1z}(x,y,z;t) &= \hat{B}_z(x,y;t)\sin(kz),
\end{split}    
\end{equation}
    where $k=n\pi/L$ is the quantized axial wavenumber $(n=1, 2, \cdots)$. 
Equations~\eqref{eq_linMHD_3Dvx} to \eqref{eq_linMHD_3DBz} then become
\begin{eqnarray}
\dfrac{\partial \hat{v}_{x}}{\partial t}
&=& -\dfrac{B_0}{\mu_0\rho_0}
  \left(k\hat{B}_{x}+\dfrac{\partial\hat{B}_{z}}{\partial x}
  \right), 
            \label{eq_linMHD_2Dvx} \\
\dfrac{\partial \hat{v}_{y}}{\partial t}
&=& -\dfrac{B_0}{\mu_0\rho_0}
  \left(k\hat{B}_{y}+\dfrac{\partial\hat{B}_{z}}{\partial y}
  \right), 
            \label{eq_linMHD_2Dvy} \\
\dfrac{\partial \hat{B}_{x}}{\partial t}
&=& k B_0 \hat{v}_{x}, 
            \label{eq_linMHD_2DBx}  \\
\dfrac{\partial \hat{B}_{y}}{\partial t}
&=& k B_0 \hat{v}_{y}, 
            \label{eq_linMHD_2DBy}  \\    
\dfrac{\partial \hat{B}_{z}}{\partial t}
&=& -B_0
    \left(\dfrac{\partial \hat{v}_{x}}{\partial x}
         +\dfrac{\partial \hat{v}_{y}}{\partial y}
    \right). 
                    \label{eq_linMHD_2DBz}
\end{eqnarray}
Axial fundamentals ($k=\pi/L$) will be examined throughout, 
    even though our analysis can be readily adapted to  
    any axial harmonic number $n$. 

\subsection{Energy Conservation Law}
\label{sec_sub_ener}
Energetics considerations turn out to be necessary.
Consider a volume $V$ that laterally occupies an arbitrary area $Q$ 
    and is axially bounded by the planes $z=0$ and $z=L$. 
Let the curve enclosing $Q$ be denoted by $\partial Q$. 
Let $[\uvec{n},\uvec{t}, \uvec{z}]$ further define a right-handed set of 
    orthonormal system at any point along $\partial Q$, with the 
    normal direction $\uvec{n}$ pointing away from $Q$.
An energy conservation law can be readily derived from
   Eqs.~\eqref{eq_linMHD_3Dvx} to \eqref{eq_linMHD_3DBz}, reading
\begin{equation}
   \dfrac{\mathd E(t)}{\mathd t}
 =-F(t), \label{eq_enercons}
\end{equation}
   where
\begin{eqnarray}
&& E(t) =  \int_{Q} \epsilon dx dy, 
                 \label{eq_defE}  \\ 
&& F(t)=  \oint_{\partial Q} dl_t
         \uvec{n}\cdot\hat{p}_{\rm T}
         \left(\hat{v}_x\uvec{x}+\hat{v}_y\uvec{y}\right),
                 \label{eq_defF} \\
&& \epsilon(x, y; t)
      = \dfrac{1}{2}\rho_0
              \left(\hat{v}_x^2+\hat{v}_y^2\right) 
        +\dfrac{1}{2\mu_0}
              \left(\hat{B}_x^2+\hat{B}_y^2+\hat{B}_z^2
              \right).
                 \label{eq_defeps}
\end{eqnarray}
Note that the ansatz~\eqref{eq_FourierAnsatz} has been employed,
    and $\hat{p}_{\rm T}=B_0\hat{B}_z/\mu_0$ is connected with the 
    Eulerian perturbation of total pressure.
Note also that $dl_t$ denotes the elementary arclength in the tangential direction
    defined by $\uvec{t}$.
When multiplied by $L/2$, the symbols $E$ and $F$ represent the instantaneous
    perturbation energy in $V$ and the net energy flux out of $V$, respectively.
We nonetheless refer to $E$ ($F$) as the total energy (the net energy flux) for brevity.
Likewise, the symbol $\epsilon$ in Eqs.~\eqref{eq_defE} and \eqref{eq_defeps} 
    will be referred to as the energy density. 
We also drop the hat from here onward.

\subsection{Equilibrium Configuration and Initial Perturbation}
\label{sec_sub_equipert}
Our equilibrium comprises a composite structure embedded in 
   an otherwise uniform external corona (denoted by subscript ${\rm e}$). 
By ``composite'' we mean two identical tubes  
   with elliptic cross-sections and separated by $d$.
Let $j$ label a tube, and let a 2D position vector $\vec{X}_j=(X_j,Y_j)$ 
   denote the center of tube $j$.  
The tube centers are placed on the $x$-axis without loss of generality, 
   enabling a tube to be referred to as either the left ($j={\rm L}$)
   or the right tube ($j={\rm R}$). 
The tube centers are further taken to be symmetric about $x=0$, 
   resulting in $X_{\rm L, R} = \mp d/2$ and $Y_{\rm L, R} = 0$.
We consistently use subscript ${\rm i}$ to denote the equilibrium quantities 
   at either tube axis, meaning in particular that the internal density
   and \Alf\ speed are denoted by $\rhoi$ and $\vAi$, respectively.  
Likewise, by $\vAe$ we denote the \Alf\ speed evaluated with the external density $\rhoe$.

Our two-tube structure is characterized as follows. 
The tubes are taken to be identical
    not only in cross-sectional shapes but in their orientations
    relative to the $x$-axis. 
For simplicity, we consider only two orientations by discriminating whether
    the major (hereafter ``$x$-major'') 
    or minor axis (``$x$-minor'') 
    of a tube aligns with the $x$-direction.
Let $a_x$ and $a_y$ denote the spatial extent that a tube spans in the 
    $x$- and $y$-directions, respectively. 
The $x$-major ($x$-minor) orientation then means that 
    $a_x>a_y$ ($a_x<a_y$).
Both orientations are realized by the density distribution
    ($j={\rm L}$ or ${\rm R}$),
\begin{eqnarray}
  \rho_j(x,y) 
= \left\{
       \begin{array}{lc}
            \rhoi,            
                            &~~~0\le \bar{r}_j\le 1-\bar{l}; \\[0.2cm]
            \dfrac{\rhoi+\rhoe}{2}
               -\dfrac{\rhoi-\rhoe}{2}
                \cos\dfrac{\pi(\bar{r}_j-1)}{\bar{l}},
                            &~~~1-\bar{l} \le \bar{r}_j\le 1,	
       \end{array}
  \right.
\label{eq_prof_rhoj}
\end{eqnarray}
    with the intermediate dimensionless variable $\bar{r}_j$ defined by
\begin{equation}
   \bar{r}_j(x,y) 
\coloneqq
   \sqrt{\left(\dfrac{x-X_j}{a_x}\right)^2
        +\left(\dfrac{y}{    a_y}\right)^2}.
\label{eq_def_rbarj}
\end{equation}
We take $d \ge 2 a_x$ such that tube $j$ can be unambiguously identified
    as where $\bar{r}_j\le 1$. 
We also use $a$ ($b$) to denote the semi-major (semi-minor) axis,
    meaning that $[a_x=a, a_y=b]$ ($[a_x=b, a_y=a]$) for
    the $x$-major ($x$-minor) orientation.
The symbol $b$ will be favored when the limiting case of circular tubes
    is examined ($a=b$).
Equation~\eqref{eq_prof_rhoj} represents a tube profile that continuously connects
    the internal density $\rhoi$ to the external one $\rhoe$ via some elliptic layer
    of width $\bar{l} a$ ($\bar{l} b$) along the direction of 
    the major (minor) axis.
When $a=b$, this profile is equivalent to the sinusoidal distribution
    introduced by \citet[][hereafter \citetalias{2002ApJ...577..475R}]{2002ApJ...577..475R} 
    for modeling circular inhomogeneities
    \citep[see e.g.,][for more applications]{2004ApJ...606.1223V,2013ApJ...777..158S,2021ApJ...908..230C}.
Somehow subtle is that the \citetalias{2002ApJ...577..475R} implementation 
    for a tube centered at $(X_j, 0)$ writes
\begin{eqnarray}
  \rho_j(x,y) =  
 \left\{
       \begin{array}{lc}
            \rhoi,            
                            &~~~                  {r}_j \le \Rtrd-\ltrd/2; \\[0.2cm]
            \dfrac{\rhoi+\rhoe}{2}
               -\dfrac{\rhoi-\rhoe}{2}
                \sin\dfrac{\pi({r}_j-\Rtrd)}{\ltrd},
                            &~~~\Rtrd-\ltrd/2 \le {r}_j \le \Rtrd+\ltrd/2, \\[0.2cm]
            \rhoe,
                            &~~~                  {r}_j \ge \Rtrd+\ltrd/2. 
       \end{array}
  \right.
\label{eq_prof_rhojRR02}
\end{eqnarray}    
Here $r_j\coloneqq\sqrt{(x-X_j)^2+y^2}$.
Evidently, the geometrical parameters $[\Rtrd,\ltrd]$ characterizing the 
    \citetalias{2002ApJ...577..475R} implementation are connected to ours
    via
\begin{equation}
\label{eq_prof_RR02toMine}
\begin{split}
& b=\Rtrd+\ltrd/2, \quad \\
& \bar{l} = \dfrac{\ltrd}{b} = \dfrac{\ltrd/\Rtrd}{1+\ltrd/2\Rtrd}.     
\end{split}
\end{equation}
    
\begin{figure*}
\centering
\includegraphics[width=.95\textwidth]{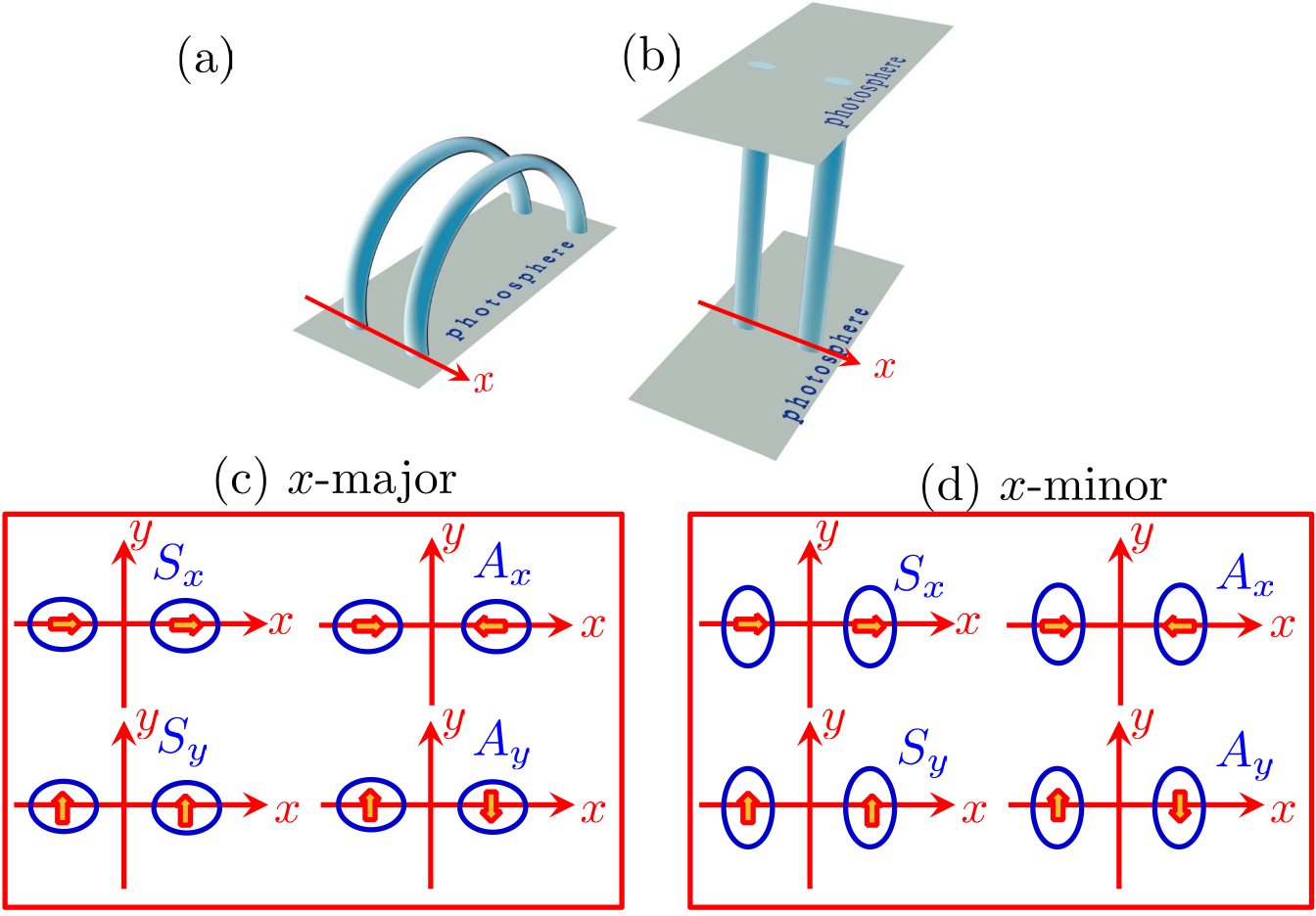}
\caption{
Schematics showing both the two-elliptic-tube configuration
    and lower-order kink motions therein.
Illustrated in (b) are two identical tubes with elliptic cross-sections
    and separated in the $x$-direction.
The $y-z$ plane is identified as the tube plane, the meaning of which 
    is clearer in the curved representation in panel a.
This study examines only two tube orientations (panels c and d), 
    with the one labeled ``$x$-major'' (``$x$-minor'')
    corresponding to the situation where the major- (minor-)axis
    of either tube is aligned with the $x$-axis. 
The tube centers are placed symmetrically with respect to the $y$-axis. 
Only axial fundamentals are considered, making it possible to examine
    the three-dimensional (3D) wave fields with a 2D approach. 
Lower-order kink motions are classified according to their 
    internal flow fields in the apex plane (panels c or d); 
    hence such symbols as $S_x$ and $A_y$. 
See text for more details. 
 }
\label{fig_Schm} 
\end{figure*}

Our equilibrium configuration, illustrated in Fig.~\ref{fig_Schm}b, 
    is a straightened version of a composite system 
    where two tubes are separated in the $x$-direction
    and anchored in the photosphere (Fig.~\ref{fig_Schm}a). 
We see the $x$- ($y$-) direction in Fig.~\ref{fig_Schm}b 
    as horizontal (vertical) given that the $y-z$ plane can be identified
    as the tube plane. 
For two-tube systems, however, it turns out to be also necessary 
    to classify lower-order kink motions into the $S_x$, $A_x$,
    $S_y$, and $A_y$ types by combining the symmetric and polarization properties 
    of the internal flow fields (see Fig.~2 in \citetalias{2008ApJ...676..717L}).
Proposed for two-identical-circular-tube systems,
    the subscript $x$ ($y$) arises when the internal velocities are dominated
    by their $x$- ($y$-)components, 
    while the symbol $S$ ($A$) is such that the dominant internal velocity component
    is symmetric (anti-symmetric) about $x=0$.
This classification is expected to 
    carry over when elliptic tubes are examined.
Somehow complicated is that one needs to distinguish between
    the $x$-major and $x$-minor arrangements (Figs.~\ref{fig_Schm}c and \ref{fig_Schm}d),
    meaning for instance that the frequencies of the $S_x$ motions 
    are likely to be different for different tube orientations. 
 
Our equilibrium is perturbed via velocities.
We consider only $S_x$ and $A_x$ motions, given that kink motions
    are commonly observed to be horizontally polarized
    (see \citealt{2023NatCo..14.5298Z} and references therein;
    see e.g., \citealt{2008A&A...489.1307W} and \citealt{2011ApJ...736..102A}
    for observational instances of vertical kink motions).
Regardless of tube orientations, these are excited by imposing  
\begin{equation}
   \dfrac{\vec{v}_{\rm ini}(x,y)}{\vAe} 
=  \Exp{4}
   \left\{
        \exp\left[-\left(\dfrac{x+x_{\rm ini}}{b}\right)^2\right]
    \pm \exp\left[-\left(\dfrac{x-x_{\rm ini}}{b}\right)^2\right] 
   \right\}
   \uvec{x},
   \label{eq_initPert}
\end{equation}
    where $x_{\rm ini}=X_{\rm R}+a_x+2b$, and we recall that
    $[X_{\rm R}=d/2, b=\min(a_x, a_y)]$.
The coefficient $\Exp{4}$ is introduced only for plotting purposes,
    the magnitude being immaterial for linear studies. 
Equation~\eqref{eq_initPert} represents
    a pair of planar compressible
    perturbations concentrated in the ambient corona,
    exciting $S_x$ ($A_x$) motions when the plus (minus) sign applies.

\subsection{Parameter Overview and Solution Method}
\label{sec_sub_solmethod}
The behavior of small-amplitude perturbations is 
    determined by two sets of parameters.
We take the dimensional set to be $\{\rhoe, b, \vAe\}$, 
   which serves merely as normalizing constants in our context.
Accept that we are primarily interested in such timescales
   as the oscillation period and some damping time. 
The importance of the dimensionless set then follows from a straightforward
   dimensional analysis, which dictates that a timescale $t_{\rm scale}$
   is formally expressible as
\begin{equation}
   \dfrac{\vAe t_{\rm scale}}{b}
 = \mathcal{F}
   \left[\mbox{orientation},
        \dfrac{\rhoi}{\rhoe}, \dfrac{a}{b}, \bar{l}, \dfrac{d}{b}; \quad 
        kb; \quad 
        \mbox{perturbation pattern}
   \right].
   \label{eq_DimAnalysis}
\end{equation} 
The subgrouped parameters characterize 
   the transverse structuring, axial wavenumber, and initial perturbation, respectively.
By ``orientation'' we refer to either ``$x$-major'' or ``$x$-minor''.
By ``perturbation pattern'' we mean either $S_x$ or $A_x$. 
We fix the axial wavenumber at $kb=\pi/30$ throughout, 
    meaning a tube length $L=30b$ for axial fundamentals.
This $L/b$ suffices for our purposes, lying toward the lower end of
    but within the accepted range for AR loops imaged in the EUV
    \citep[e.g.,][]{2004ApJ...600..458A,2007ApJ...662L.119S}.
Table~\ref{tab_solset} briefly overviews our computations,
    which are grouped into four sets to be detailed later.
The density contrast $\rhoi/\rhoe$ is fixed at either $3$ or $5$, 
    \revise{both values being representative of AR loops
    \citep[e.g.,][and references therein]{2004ApJ...600..458A}.}

We solve Eqs.~\eqref{eq_linMHD_2Dvx} to \eqref{eq_linMHD_2DBz}
   with the following procedure.
A computational domain $[-\xM, \xM]\times [-\yM, \yM]$
   is discretized into a uniform mesh with identical spacing
   in the $x$- and $y$-directions
   ($\Delta_x=\Delta_y=\Delta$).
The zero-gradient condition is implemented for all unknowns
   at all boundaries.
We evolve Eqs.~\eqref{eq_linMHD_2Dvx} to \eqref{eq_linMHD_2DBz}
   with the classic MacCormack scheme \citep{1969AIAA..354Mac},
   a popular finite-difference algorithm second-order accurate
   in both space and time 
   \citep[see the textbooks by, e.g.,][for more]{1995Anderson_CFD,jardin2010computational}.
The time step $\Delta_t$ is set via some effective Courant number
   $c\coloneqq (\vAe\Delta_t/\Delta_x)^{2/3}+(\vAe\Delta_t/\Delta_y)^{2/3}$
   as inspired by \citet{1996ApplMathLett...9..99H}.
A value of $c=0.99$ is employed for all the presented results.
We have verified that varying $c$ between $\sim 0.7$ and $\sim 1.3$
   introduces no discernible difference, and numerical stability
   is consistently maintained.

Some remarks on the grid setup are necessary. 
We start by noting that the spatial spacing $\Delta$ 
   restricts the timeframe in which numerical solutions make physical sense,
   an aspect raised by 
   \citet[][hereafter \citetalias{2008ApJ...679.1611T}]{2008ApJ...679.1611T}
   who numerically solved an equivalent set of governing equations. 
Defining the \Alf\ frequency $\omgA(x, y) = k \vA(x, y)$,
   one expects that some time-dependent phase-mixing length will emerge as
   \citep{1995JGR...10019441M}
\begin{equation}
   \Lph(x, y; t) = \dfrac{2\pi}{|\nabla\omgA(x,y)| t}, 
\end{equation}
   which characterizes the transverse lengthscales of resonantly generated
   \Alfvenic\ motions in the nonuniform portions in the system.
It suffices to consider only the nonuniform layer of one tube.
Evidently, the shortest phase-mixing length $\Lph^{\rm min}(t)$ 
   at a given instant occurs at the strongest $|\nabla\omgA(x,y)|$.
Equally evident is that this strongest $|\nabla\omgA(x,y)|$
   depends only on $\rhoi/\rhoe$ and $\bar{l}$ even when elliptic tubes
   are examined.
Let $\Lph^{\rm min}(500)$ denote $\Lph^{\rm min}$ at $t=500b/\vAe$,
   before which our computations are consistently terminated.
It then follows from the arguments by \citetalias{2008ApJ...679.1611T}
   that our signals are physically relevant provided $\Lph^{\rm min}(500)\ge 2\Delta$.  
As shown by Table~\ref{tab_solset}, 
   this criterion is satisfied by the reference grid setup
   \revise{for any set of our computations}.

\begin{table*}[h!]
\caption{Summary of time-dependent solutions presented in the text}
\label{tab_solset}
\centering
\begin{tabular}{l||l|l|l||l||l|r}
\hline
Set    & $\rho_{\rm i}/\rho_{\rm e}$     & $\bar{l}$         & $\Lph^{\rm min}(500)$ 
       & Reference Grid     
       & Section                          & Remarks \\ 
\hline
1      & $3$                             & $\ge 0.2$       & $\gtrsim 0.032b$           
       & $[\xM, \yM]=[12b, 12b], \Delta=0.01b$         
       & \S\ref{sec_sub_1circ}           & \makecell{1-circular-tube}          \\ 
\hline
2      & $5$                             & $\gtrsim 0.182$ & $\gtrsim 0.019b$                        
       & $[\xM, \yM]=[12b, 10b], \Delta=0.005b$  
       & \S\ref{sec_sub_2circ}           &\makecell{2-circular-tube}\\ 
\hline
3      & $3$                             & $0.4$           & $0.063b$ 
       & $[\xM, \yM]=[30b, 25b], \Delta=0.01b$  
       & \S\ref{sec_sub_2ellip_xMj}      & \makecell[l]{2-elliptic-tube\\$x$-major} \\ 
\hline
4      & $3$                             & $0.4$           & $0.063b$
       & $[\xM, \yM]=[30b, 25b], \Delta=0.01b$
       & \S\ref{sec_sub_2ellip_xMn}      & \makecell[l]{2-elliptic-tube\\$x$-minor} \\ 
\hline
\end{tabular}
\tablefoot{All computations pertain to $kb=\pi/30$ or equivalently $L/b=30$
    for axial fundamentals.}
\end{table*}

\section{Test Computations for Circular Tubes}
\label{sec_verif}
The MacCormack scheme, while a textbook one, has not been applied to
   Eqs.~\eqref{eq_linMHD_2Dvx} to \eqref{eq_linMHD_2DBz}.
Its applicability is therefore examined in this section via some
   test computations for which the time-dependent behavior
   can be established or expected with independent methods.   
Only circular tubes are of interest, and we refer to $b$($=a$)
   as our tube radius. 
We start by examining the response of a one-circular-tube system to
   axisymmetric (sausage) perturbations, following our previous study \citep{2022ApJ...928...33L} to formulate 
   both the equilibrium and the initial perturbation. 
The Fourier-integral-based solutions, presented in Fig.~4 therein, agree
   closely with our MacCormack solutions.
In particular, no discernible numerical anisotropy 
   shows up even though finite differences are performed on a Cartesian grid
   to examine a non-planar equilibrium 
   \citep[see e.g., the review by][for more on numerical anisotropy]
   {2015AdvDiffEq...9..1}.
The rest of this section focuses on kink perturbations. 

\subsection{Kink Motions in a One-Circular-Tube Setup}
\label{sec_sub_1circ}
This subsection examines kink perturbations in a one-circular-tube configuration,
    placing the tube center at the origin without loss of generality.
The equilibrium density is realized through
    Eqs.~\eqref{eq_prof_rhoj} and \eqref{eq_def_rbarj}
    by taking $[a_x=a_y=b, X_j=0]$ and then dropping the subscript $j$.
Kink motions in such a configuration have been extensively studied
    \citep[see the review by][and references therein]{2021SSRv..217...73N},
    readily allowing our MacCormack computations to be compared with known results.
By ``known'' we refer to two sets of studies. 
Set one, presented in Sect.~4 of \citetalias{2008ApJ...679.1611T},
    adopts the \citetalias{2002ApJ...577..475R} implementation to examine
    the system response to an initial perturbation of the form
\begin{equation}
   \vec{v}_{\rm ini}(x,y) 
=  \vAe  
   \exp\left[-\dfrac{(y-y_{\rm ini})^2}{\sigma^2}\right]
   \uvec{y}.
   \label{eq_initPertT08}    
\end{equation}
Figure~2 in \citetalias{2008ApJ...679.1611T} then presents the time sequence
    of $v_y$ sampled at the tube center for a combination of parameters
    $[\rhoi/\rhoe=3, k\Rtrd=\pi/20, \ell/\Rtrd=0.6]$ 
    and $[y_{\rm ini}=3\Rtrd, \sigma=\Rtrd]$.
We repeat the same experiment, using Eq.~\eqref{eq_prof_RR02toMine}
    to address notational differences and adopting an identical computational grid. 
Our finite-difference results are found to be consistent with 
    \citetalias{2008ApJ...679.1611T}, despite that the numerical code
    therein adopts the finite-volume methodology.      

We proceed to perform an additional series of computations to examine
    whether our code outputs agree with expectations
    for ideal kink quasi-modes. 
Let $\omega=\Omega-i\gamma$ denote the complex-valued quasi-mode frequency,
    and see $\Omega$ and $\gamma$ as positive.
With the kink speed $\ck$ defined as
\begin{equation}
    \ck^2 \coloneqq \dfrac{\rhoi\vAi^2+\rhoe\vAe^2}{\rhoi+\rhoe},
\end{equation}
   it is well established that (\citetalias{2002ApJ...577..475R}; see also
    \citealt{1992SoPh..138..233G} and \citealt{2013ApJ...777..158S})
\begin{equation}
\begin{split}
   & \Omega \approx k \ck, \\[0.2cm] 
   & \dfrac{\gamma}{\Omega}
     \approx 
     \dfrac{1}{4}
     \dfrac{\bar{l}}{1-\bar{l}/2}
     \dfrac{\rhoi/\rhoe-1}{\rhoi/\rhoe+1}
\end{split}
\label{eq_expectTTTB}
\end{equation}
    in the so-named thin-tube-thin-boundary limit
    (TTTB, $kb \ll 1$ and $\bar{l}\ll 1$).
Note that the TTTB expressions are usually formulated in 
    terms of the \citetalias{2002ApJ...577..475R} notations $[\Rtrd,\ltrd]$.
Note also that the TTTB results may not hold well beyond their range
    of applicability \citep[e.g.,][]{2014ApJ...781..111S,2021SoPh..296...95Y}.
We therefore also evaluate the mode frequencies numerically with  
    the general-purpose finite-element code PDE2D \citep{1988Sewell_PDE2D},
    computing ideal kink quasi-modes as resistive eigenmodes
    (see \citealt{2011SSRv..158..289G} for conceptual clarifications;
    see also \citealt{2005ApJ...618L.149T} for the first introduction
    of PDE2D to solar contexts)
    \footnote{A uniform resistivity $\eta$ is employed for simplicity, the associated magnetic Reynolds number $R_{\rm m}\coloneqq \mu_0 \Rtrd \vAi/\eta = 10^5$.
   We have verified that further increasing $R_{\rm m}$ does not introduce any discernible difference to the oscillation frequencies and damping rates. 
   These $\eta$-independent values are well known to be physically connected to ideal quasi-modes \citep[e.g.,][]{1991PhRvL..66.2871P,2006ApJ...642..533T,2013ApJ...777..158S}.}.
No restriction is necessary for $kb$ or $\bar{l}$.

\begin{figure*}
\centering
\includegraphics[width=.85\textwidth]{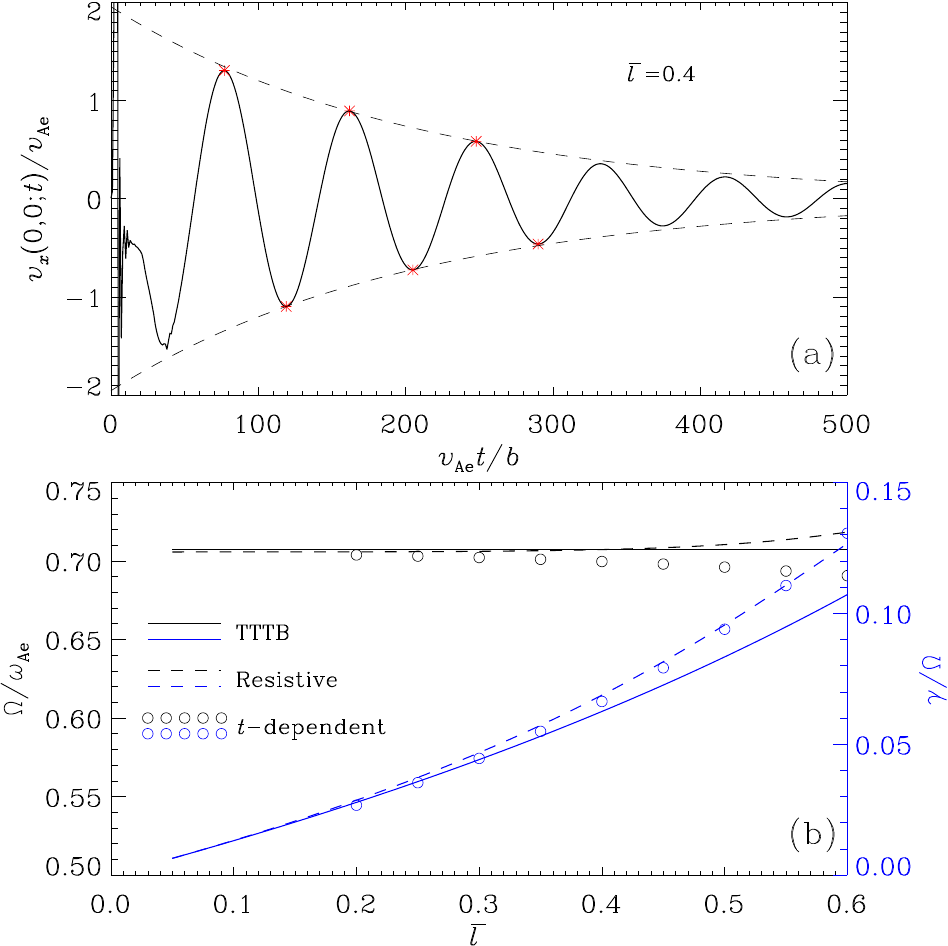}
\caption{Kink motions in isolated tubes with circular cross-sections
    for a fixed pair of density contrast and dimensionless axial wavenumber
    $[\rhoi/\rhoe=3, kb=\pi/30]$. 
A specific dimensionless layer width $\bar{l}=0.4$
    is examined in (a), the solid curve showing the $x$-speed
    sampled at the tube center ($v_x(0,0,t)$).
An exponentially damped cosine is employed to fit the segment encompassing
    the extrema shown by the red asterisks, 
    with the resulting damping envelope plotted over the entire interval 
    (the dashed curves).
Presented in (b) are the $\bar{l}$-dependencies of
    the mode frequencies ($\Omega$, the black curves and symbols)
    and the ratios of the decay rate to the frequency ($\gamma/\Omega$, blue)
    obtained with various methods. 
The solid curves represent the analytical thin-tube-thin-boundary (TTTB) expectations
    with Eq.~\eqref{eq_expectTTTB}.
No assumption on tube length or layer width is imposed 
    in the ``Resistive'' computations (the dashed curves) 
    that evaluate ideal quasi-modes as resistive eigenmodes. 
Shown by the open circles are the best-fit results from our time-dependent computations,
    where kink motions are consistently excited with a perturbation
    given by Eq.~\eqref{eq_initPer1Loop}.
 }
\label{fig_1tubekink} 
\end{figure*}

The rest of this subsection is devoted to a fixed combination
   $[\rhoi/\rhoe=3, kb=\pi/30]$, allowing only $\bar{l}$ to vary. 
Furthermore, the reference grid (see Table~\ref{tab_solset})
   is consistently employed in our time-dependent computations,
   where kink motions are excited by a perturbation of the fixed form 
\begin{equation}
   \dfrac{\vec{v}_{\rm ini}(x,y)}{\vAe} 
=  \Exp{4}  
   \exp\left[-\dfrac{(x+3b)^2}{b^2}\right]
   \uvec{x}.
   \label{eq_initPer1Loop}    
\end{equation}
Equation~\eqref{eq_initPer1Loop} is essentially identical to 
    Eq.~\eqref{eq_initPertT08}
    except that $v_x$ rather than $v_y$ is perturbed. 
Figure~\ref{fig_1tubekink}a examines a representative case with $\bar{l}=0.4$,
    showing the temporal evolution of the $x$-speed sampled at the tube center
    ($v_x(0,0,t)$, the solid curve). 
This $v_x$ signal is seen to feature some rapid variations for $t\lesssim 15b/\vAe$. 
By ``rapid'' we mean timescales on the order of the transverse \Alf\ time ($b/\vAi$), 
    which derives from the multiple reflections off the tube boundaries of the perturbations
    that are imparted by the external driver to the tube. 
The $v_x$ signal transitions toward a regular slower-varying pattern afterwards, 
    becoming monochromatic when $t\gtrsim 70b/\vAe$. 
We choose to leave out the first extremum in this stage for safety,
    and fit the segment encompassing the next six (the red asterisks) 
    with an exponentially damped cosine
\begin{equation}
    f_{\rm fit} (t) \propto 
      \exp{\left(-\dfrac{t}{\tau}\right)}
      \cos\left(\dfrac{2\pi t}{P}+\phi\right)
   \label{eq_funcfit}.
\end{equation}
The damping envelope from the best-fit
    is then plotted for the entire duration in Fig.~\ref{fig_1tubekink}a
    by the dashed curves. 
One see that this best-fit envelope, while deduced for some segment,
    offers a good description for the larger-time behavior    
    as well (say, $t\gtrsim 300b/\vAe$).

Figure~\ref{fig_1tubekink}b surveys a range of $\bar{l}$ 
    by plotting the oscillation frequencies ($\Omgfit$, the black open circles)
    and damping rates ($\gamfit$, blue), which are translated from
    the best-fit periods and damping times
    ($\Omgfit = 2\pi/\Pfit$ and $\gamfit=1/\taufit$).
Note that $\Omega$ is measured in units of the external \Alf\ frequency 
    $\omgAe=k\vAe$,
    and $\gamma/\Omega$ is presented rather than $\gamma$ itself. 
The curves in Fig.~\ref{fig_1tubekink}b further provide the quasi-mode expectations
    with either the analytical TTTB expression (Eq.~\ref{eq_expectTTTB}, the solid lines)
    or the PDE2D computations (labeled ``Resistive'', dashed).
One sees that the TTTB results provide a rather good approximation to 
    the numerical ``Resistive'' ones, the frequencies being practically the same 
    and the damping rates differing by $\gtrsim 10\%$ only when $\bar{l}\gtrsim 0.4$. 
One further sees that the open circles agree well with the ``Resistive'' 
    computations, demonstrating that the internal flow fields practically 
    evolve as an ideal quasi-mode at least in the interval where 
    the fitting is performed.

\subsection{Kink Motions in a Two-Circular-Tube Setup}
\label{sec_sub_2circ}
This subsection examines kink motions in a two-circular-tube configuration, 
    for which the distinction between $S_x$ and $A_x$ patterns becomes necessary. 
The initial perturbation is therefore chosen to follow Eq.~\eqref{eq_initPert},
    where $x_{\rm ini}$ evaluates to $d/2+3b$. 
Our code consistently adopts the $[b, \bar{l}]$ notations.
However, the mixed usage of the \citetalias{2002ApJ...577..475R} convention
    $[\Rtrd, \ltrd]$ turns out to be necessary.
The reason is that this \citetalias{2002ApJ...577..475R} convention was adopted 
    by \citetalias{2015A&A...582A.120S}
    to examine ideal quasi-modes from an eigenvalue-problem perspective
    in a configuration physically identical to ours.
Briefly put, the study by \citetalias{2015A&A...582A.120S} is based on the
    TB-embedded T-matrix formalism of scattering theory, 
    the pertinent results being most straightforward
    for our time-dependent computations to be compared with. 
\revise{By ``pertinent'' we specifically} refer to Fig.~4 therein, which presents 
    the $\ltrd/\Rtrd$-dependencies of the quasi-mode frequencies and damping rates
    for a fixed combination $[\rhoi/\rhoe=5, d/\Rtrd=2.5, k\Rtrd=\pi/100]$.
Note that varying $\ltrd/\Rtrd$ actually impacts our $\bar{l}$, $d/b$,
    and $kb$ simultaneously (see Eq.~\ref{eq_prof_RR02toMine}). 
We adopt a fixed $kb=\pi/30$ to save computational time. 
This does not matter because we will measure our timescales or frequencies 
    in appropriate units (say, $\Omega$ in $\omgAe$) and the resulting
    readings do not depend on $k\Rtrd$ when $k\Rtrd\lesssim \pi/20$
    (see Fig.~3 in \citetalias{2015A&A...582A.120S}, and note that
    $k\Rtrd < kb$).
Equation~\eqref{eq_prof_RR02toMine} readily converts the pair
    $[\ltrd/\Rtrd, d/\Rtrd]$ into our $[\bar{l}, d/b]$. 

\begin{figure*}
\centering
\includegraphics[width=.85\textwidth]{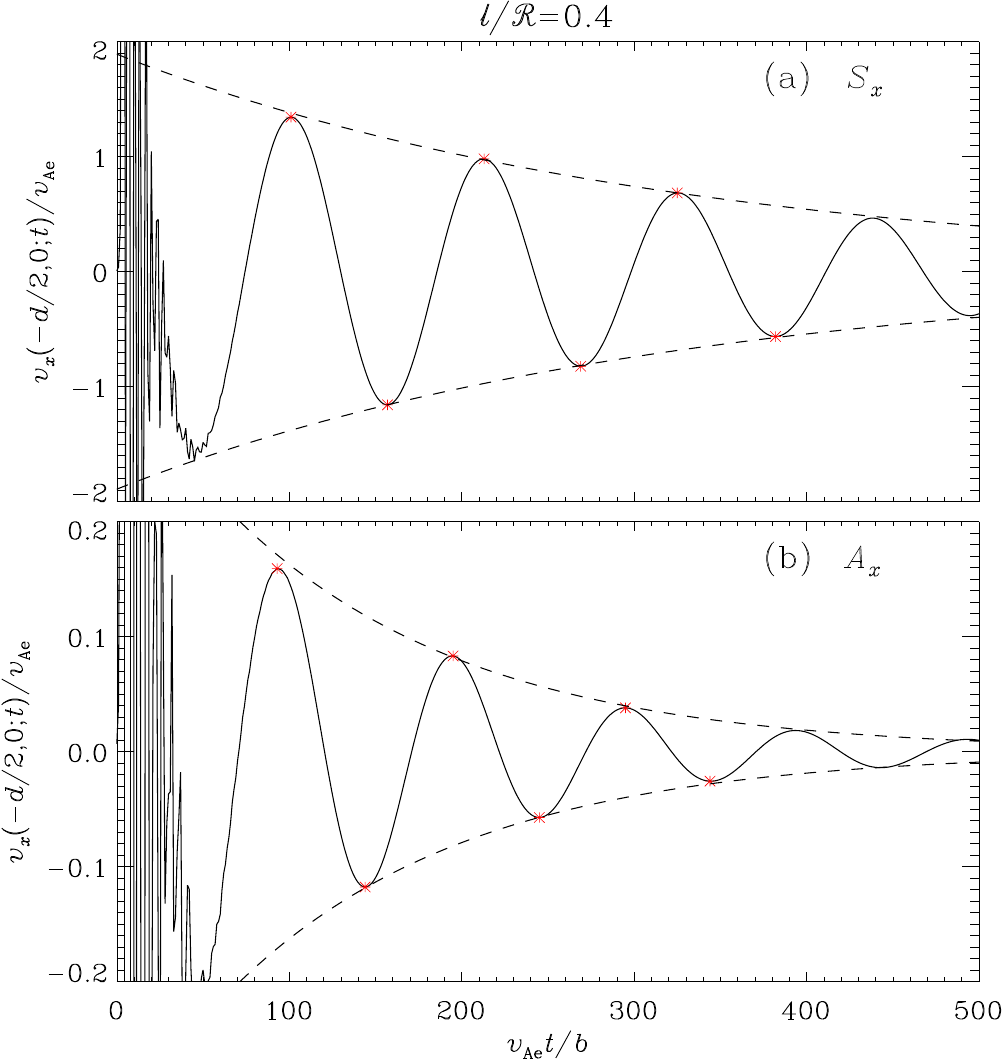}
\caption{
Kink motions with (a) $S_x$ and (b) $A_x$ patterns in 
   a two-circular-tube configuration.
Plotted are the time sequences of the $x$-speed sampled at the left tube center
   ($v_x(-d/2, 0;t)$, the solid curves) together with
   the best-fit damping envelopes (dashed). 
The time-dependent computations pertain to a fixed combination
   $[\rhoi/\rhoe=5, \ltrd/\Rtrd=0.4, d/\Rtrd=2.5, kb=\pi/30]$, with $\ltrd$ and $\Rtrd$
   being the nonuniform layer width and mean tube radius 
   implemented by \citet{2002ApJ...577..475R}.
A fitting procedure with an exponentially damped cosine 
   is performed over the duration exactly encompassing
   the extrema indicated by the red asterisks, even though the best-fit envelopes
   are plotted for the entire time interval.
Kink motions are excited with the perturbation described by Eq.~\eqref{eq_initPert}
   with appropriate signs. 
 }
\label{fig_2tubeTst_SAfit} 
\end{figure*}

\revise{We now test}
    our time-dependent computations against 
    the \citetalias{2015A&A...582A.120S} results, 
    to be labeled ``T-matrix'' for clarity.
It suffices to examine the time sequences of the $x$-speed sampled
    at the left tube center (namely, $v_x(-d/2, 0; t)$);
    the right counterpart is either identical (for $S_x$)
    or different only by sign (for $A_x$). 
Figure~\ref{fig_2tubeTst_SAfit} focuses on the choice $\ltrd/\Rtrd=0.4$,
    and plots the sampled $v_x$ for (a) the $S_x$ and (b) the $A_x$ patterns
    by the solid curves. 
The same analysis as in Fig.~\ref{fig_1tubekink}a is then repeated for both curves, 
    whereby we single out the six extrema (the red asterisks in Fig.~\ref{fig_2tubeTst_SAfit})
    after the transitory phase to perform a fitting with Eq.~\eqref{eq_funcfit}. 
\revise{The best-fit exponential envelope, plotted by the dashed curves for the entire duration,
    is seen to well reflect the damping of kink motions for much longer time.}

Figure~\ref{fig_2tubeTst_cpTMatrix} proceeds to examine 
    the $\ltrd/\Rtrd$-dependencies 
    of (a) the oscillation frequencies ($\Omega$ in $\omgAe$)
    and (b) damping-rate-to-frequency ratios ($\gamma/\Omega$).
The $S_x$ and $A_x$ patterns are discriminated by the different colors.
We present the best-fit values from our time-dependent computations 
    by the open circles, and overplot the T-matrix results by the solid curves
    for comparison. 
Note that these T-matrix curves are read from Fig.~4 
    of \citetalias{2015A&A...582A.120S}.
Our best-fit results are seen to compare favorably with the T-matrix results,
    which is particularly true when $\ltrd/\Rtrd\lesssim 0.3$.
This statement holds despite the somewhat visible difference 
    in $\Omega$ at, say, $\ltrd/\Rtrd=0.2$,
    where the best-fit value actually deviates from its T-matrix counterpart
    by only $\sim 2.7\%$. 
One further sees that the most significant departure
    occurs for the $A_x$ damping rate when $\ltrd/\Rtrd=0.4$.
However, this difference is rather modest and reads $\sim 21.7\%$
    in relative terms. 
We therefore conclude that the dispersive properties of kink quasi-modes 
    can be reasonably computed by incorporating the shortcut TB formulae
    in the T-matrix framework for all $\ltrd/\Rtrd$ examined here
    \footnote{Some discrepancy exists in the literature regarding the damping of lower-order kink motions when the tubes are in contact. 
    The T-matrix results by \citetalias{2015A&A...582A.120S} indicate
        that these motions remain damped (see Fig.~2 therein), whereas damping was found to disappear in the modal studies formulated in bi-cylindrical coordinates
        \citep{2011A&A...525A...4R,2014A&A...562A..38G}.
    Our IVP study supports the \citetalias{2015A&A...582A.120S} conclusion.
    The use of a bi-cylindrical coordinate system was recognized by the referenced studies
        to be physically problematic for small tube separations.}.
Our time-dependent results, on the other hand, further corroborate 
    the \citetalias{2015A&A...582A.120S} conclusion that
    $A_x$ motions posses higher frequencies
    and damp more rapidly than $S_x$ ones. 

\subsection{Interim Summary on Numerical Aspects}
\label{sec_sub_NumAspects}
\revise{This subsection further justifies our numerical treatment.
An additional series of computations have been performed 
    to address the effects of the grid spacing $\Delta$
    and the domain size $[\xM, \yM]$.
These results are collected in Appendix~\ref{sec_App_num} to streamline the main text.}
Overall, we have verified that the MacCormack scheme is appropriate for our purposes,
    with no issue arising from the application to nonplanar structures
    of the finite-difference methodology on a Cartesian grid. 
The zero-gradient boundary condition is somehow not fully transparent to perturbations
    excited by our planar drivers, and the domain size somehow impacts how
    the internal flows transition to a quasi-mode behavior.
However, this domain size effect is practically negligible
    on the best-fit periods ($\Pfit$) and damping times ($\taufit$)
    that we derive for the quasi-mode stage.
Likewise, the computed internal flows are not affected by the grid spacing $\Delta$
    provided that $\Delta$ is sufficiently small.

\section{Kink Motions in a Two-Elliptic-Tube Setup}
\label{sec_2ellip}

This section is devoted to kink motions in a two-elliptic-tube configuration, 
    for which the $x$-major and $x$-minor orientations need to be discriminated. 
The reference grid ($[\xM=30b, \yM=25b, \Delta=0.01b]$, see Table~\ref{tab_solset})
    will be consistently adopted, enabling a meaningful comparison between
    \revise{the tube orientations}. 
The notations $[a, b, \bar{l}]$ are adopted throughout this section,
    where all computations pertain to    
    a fixed combination of physical parameters $[\rhoi/\rhoe=3, \bar{l}=0.4, kb=\pi/30]$.
We adjust only the dimensionless tube separation ($d/b$) and 
    the ratio of the semi-major to semi-minor axis ($a/b$) for
    a given orientation and a given perturbation pattern 
    (see Eq.~\ref{eq_DimAnalysis}).
\revise{Tube overlapping is always avoided.} 
The equilibria are consistently perturbed with Eq.~\eqref{eq_initPert}.    

\subsection{$S_x$ and $A_x$ Motions for the $x$-major Orientation}
\label{sec_sub_2ellip_xMj}
This subsection addresses the $x$-major orientation.
We start with Fig.~\ref{fig_xmajor_vxt} to present the temporal evolution 
    of the $x$-speed sampled at the left tube center
    ($v_x(-d/2, 0; t)$, the solid lines)
    for both (a) the $S_x$ and (b) the $A_x$ motions.
The tube separation is fixed at $d=5b$, whereas a number of values
    are examined for $a/b$ as discriminated by the different colors. 
Only the variations after the rapid-varying phase are of interest
    for any $v_x$ sequence.
We somehow emphasize the segment exactly encompassing
    the six extrema starting with the one labeled by the relevant asterisk, 
    performing a fitting procedure with Eq.~\eqref{eq_funcfit}. 
A best-fit exponential envelope results, with an example plotted over 
    the entire duration for the case where $a/b=2.5$. 
With this example we demonstrate that an exponentially damped cosine
    is in general adequate for describing the $v_x$ sequences 
    after the rapid-varying phase.
The best-fit periods and damping times thus derived will be 
    be understood as pertaining to an ideal quasi-mode. 

The temporal attenuation of the internal flow fields means energy redistribution
    in ideal MHD. 
Let us take the case with $a/b=2.5$ in Fig.~\ref{fig_xmajor_vxt},
    and note that the two tubes 
    are actually in contact ($d=5b=2a$).
Figure~\ref{fig_xmajor_2dfield} presents the spatial distribution of 
    the velocity fields ($\vec{v}=v_x \uvec{x}+v_y\uvec{y}$, the blue arrows) 
    and the instantaneous energy density ($\epsilon$, filled contours)
    for (a) the $S_x$ and (b) $A_x$ motions
    at an arbitrarily chosen instant $t=260b/\vAe$.
A subarea of the second quadrant is singled out by each inset
    to emphasize the $\epsilon$ distributions in the nonuniform layers
    of both tubes, with the red curves delineating the layer boundaries. 
\revise{The white curve further indicates where the local \Alf\ frequency equals
    the quasi-mode frequency.}
Figure~\ref{fig_xmajor_2dfield} is actually taken from the attached animation.
The arrows and filled contours are plotted in a way that it makes sense
    to compare the $\vec{v}$ or $\epsilon$ strengths 
    not only
    in one snapshot but between different instants for a given orientation.   
In particular, the darker a portion is, 
    the larger the value of $\epsilon$ therein. 

Consider first the ambient fields for both perturbation patterns, 
    where by ``ambient'' we refer to the flows some distance away from the tubes.  
Likewise, by ``tube flow'' we mean the field excluding the ambient flow,
    and by ``internal flow'' we refer specifically
    to the uniform portion inside either tube.
For both patterns, one sees from the animation 
    that the ambient flows tend to vary more rapidly than the internal ones. 
This feature is particularly clear for the $S_x$ pattern because the ambient flows 
    tend to be substantially stronger than the internal flows as well.
Regardless, a periodogram analysis yields that the ambient flows follow primarily 
    the external \Alf\ frequency $\omgAe$, whereas the variations of the internal flows 
    are by far dominated by the lower quasi-mode frequency. 
That the ambient motions are not coordinated with the internal ones is a result
    of the form of the initial exciter given by Eq.~\eqref{eq_initPert}.
The relevant physics is actually very similar to what happens
    in the \citetalias{2008ApJ...679.1611T} study despite the configurational differences,
    the key being the $y$-invariance of Eq.~\eqref{eq_initPert}.   
It proves easier to explain this by considering a uniform equilibrium with density $\rhoe$, 
    for which Eqs.~\eqref{eq_linMHD_2Dvx} to \eqref{eq_linMHD_2DBz} can be combined 
    to yield a Klein-Gordon equation if the $y$-dependence is dropped
    \citep[see Eq.~1 in][]{2005ApJ...618L.149T}.
Suppose that an individual component in Eq.~\eqref{eq_initPert} is applied.
A dispersive fast wave then results, whereby any fluid parcel in the system eventually
    ends up in some oscillatory wake at the \Alf\ frequency $\omgAe$
    \citep[see Fig.~1 in][]{2005ApJ...618L.149T}.
Oscillatory wakes turn out to still exist in the ambient flow when the two elliptic tubes
    are introduced, and when an additional planar perturbation is implemented.
Note that the initial perturbations are external to our tubes.
Note further that overall the large-time behavior for the ambient flow actually  
    comprises two component wakes, each being
    the response to the corresponding component driver.
The two component wakes tend to interfere constructively (largely destructively)
    for the $S_x$ ($A_x$) pattern, thereby explaining the strength of the ambient flow
    relative to the internal one. 
    
Now examine the tube flows.
Focusing on the blue arrows, one sees from Fig.~\ref{fig_xmajor_2dfield} that 
    the gross patterns for both $S_x$ and $A_x$ are similar to 
    Fig.~2 in \citetalias{2008ApJ...676..717L} despite that a two-circular-tube equilibrium
    with piece-wise constant profile was addressed therein. 
Specifically, the internal velocity field is more or less uniform,
    and a pair of vortical motions develop at the edge of either tube. 
These two features, combined with the overall symmetric properties 
    of the internal field about $x=0$,
    demonstrate the robustness of the classification scheme for lower-order kink motions.
The primary difference from \citetalias{2008ApJ...676..717L}, on the other hand,
    is that the vortical motions gradually evolve from a simple dipolar behavior
    (see the first several instants in the animation)
    into multiple shearing layers (Fig.~\ref{fig_xmajor_2dfield}). 
This evolution is well known for kink motions in isolated tubes with
    either circular \citep[e.g.,][]{2008ApJ...687L.115T,2010ApJ...711..990P,2014ApJ...787L..22A}
    or elliptic cross-sections 
    \citep[e.g.,][]{2003A&A...409..287R,2020ApJ...904..116G}.
What Fig.~\ref{fig_xmajor_2dfield} demonstrates is then that tube interactions do not
    compromise the phase-mixing physics behind this evolution.      
That said, tube interactions do impact the detailed development of
    the small-scale shears,
    as evidenced by the differences in the velocity fields for the two different 
    perturbation patterns.
Suppose that tube interactions are negligible.
The tube flows are then expected to be identical 
    for the $S_x$ and $A_x$ patterns, meaning in particular a symmetric distribution of 
    the energy density ($\epsilon$)
    with respect to the minor axis ($x=-d/2=-2.5b$ here). 
However, this expectation roughly holds only for the $S_x$ pattern 
    (Fig.~\ref{fig_xmajor_2dfield}a inset), whereas 
    the left portion in the left tube 
    is favored in the $\epsilon$ distribution for the $A_x$ pattern
    (Fig.~\ref{fig_xmajor_2dfield}b inset). 
Besides phase-mixing, a closely related aspect that is not fundamentally compromised
    by tube interactions is the resonant interplay between the quasi-mode 
    and the \Alf\ continuum
    (see \citealt{2015ApJ...803...43S} and references therein
    for the subtle distinction between resonant absorption and phase-mixing). 
By this we refer to two features common to the $S_x$ and $A_x$ patterns,
    and we concentrate on the insets in the animation.
Firstly, the attenuation of the internal field is accompanied by the accumulation 
    of perturbation energies in the nonuniform layer.
Secondly, this energy accumulation leads to a localized $\epsilon$ distribution,
    with the strongest perturbations tending to the white resonance contour   
    as time proceeds for roughly one quasi-mode period after the rapid-varying phase.
Overall with Fig.~\ref{fig_xmajor_2dfield} we conclude that 
    the notions of phase-mixing and resonant absorption remain applicable to
     kink motions in our two-elliptic-tube configuration. 

Figure~\ref{fig_xmajor_yProf} further examines the phase-mixing process 
    by showing the $y$-cuts of the $x$-speeds through the left tube center
    for both (a) the $S_x$ and (b) the $A_x$ patterns. 
A number of instants are rather arbitrarily chosen and 
    discriminated by the different colors. 
The vertical dotted lines mark the borders of the nonuniform layer, 
    where shearing motions ($\partial v_x/\partial y$ here) with
    increasingly small scales are seen to develop as time proceeds.
We choose to quantify this development 
    by relating the number of the $v_x$ extrema ($\Nextrm(t)$) 
    to the instantaneous phase variation accumulated 
    over the \Alf\ continuum 
    ($\phiac(t)$, see the discussion on Eq.~\ref{eq_vxMann}).
Note that $\phiac(t)=(\omgAe-\omgAi)t$ is identical for 
    both perturbation patterns, corresponding specifically to $[2.82, 4.23, 5.64]$
    half-cycles for the examined instants. 
One then expects $[2^{+1}_{+0}, 4^{+1}_{+0}, 5^{+1}_{+0}]$ extrema
    in the $v_x$ profiles in Figs.~\ref{fig_xmajor_yProf}a and \ref{fig_xmajor_yProf}b,
    which is indeed the case
    \footnote{The velocity shears are prone to the Kelvin-Helmholtz instability (KHi), as was first theoretically recognized in the wave context by \citet{1983A&A...117..220H} and \citet{1984A&A...131..283B}.
    Further nonlinear simulations have demonstrated that the KHi may arise in the early stage of the evolution of kink motions for isolated tubes \citep[e.g.,][]{2008ApJ...687L.115T,2015ApJ...809...72A,2021ApJ...908..233S} and multi-tube setups alike \citep[e.g.,][]{2016ApJ...823...82M,2019ApJ...883...20G}.
    The applicability of such linear results as presented here may therefore need to be assessed on a case-by-case basis when real data are analyzed \citep[see Sect.~10 of the review by][for more on the nonlinear aspects]{2021SSRv..217...73N}.}. 
As can be readily verified, 
    that $\Nextrm(t) = \lfloor\phiac(t)/\pi\rfloor$ at these instants
    actually constrains the phase $\varphi_0$ in Eq.~\eqref{eq_vxMann}
    to a very narrow range $[q\pi, (q+0.19)\pi]$ with $q=0, 1$.

Figure~\ref{fig_xmajor_survey} proceeds to examine 
    (a) the periods $P$ and (b) damping-time-to-period ratios $\tau/P$
    of kink quasi-modes pertaining to both 
    the $S_x$ (the solid curves) and the $A_x$ (dashed) patterns, 
    showing how $P$ and $\tau$ depend on the ratio of
    the semi-major to semi-minor radius ($a/b$) 
    for a given tube separation ($d$).
We also examine a number of $d$ as discriminated by the different colors.
Note that this survey adopts a fixed combination of physical parameters
    $[\rhoi/\rhoe=3, \bar{l}=0.4, kb=\pi/30]$.
The same set of parameters is additionally adopted to evaluate $P$ and $\tau/P$
    of the kink quasi-mode for an isolated circular tube 
    with the resistive eigenmode approach (see Sect.~\ref{sec_sub_1circ}), 
    the results being plotted by the horizontal lines for comparison.
Note that the values for 
    $P$ (in units of the axial \Alf\ time 
    $\Taxial = 2\pi/\omgAe = 2L/\vAe$)
    and $\tau/P$ vary little if $kb = \pi b/L$ is further reduced,
    meaning that little will change if one examines axial fundamentals in
    AR loops with much larger $L/b$ than adopted here.
Note further that $a/b$ is required not to exceed $d/2b$ for a given $d/b$
    such that tube overlapping is avoided. 
For any given pair $[a/b, d/b]$,     
    Fig.~\ref{fig_xmajor_survey} indicates that the $A_x$ motion
    possesses a shorter period and damps more efficiently than the $S_x$ motion.
This behavior is identical to what \citetalias{2015A&A...582A.120S}
    found for two-circular-tube configurations.
The argument therein is that $A_x$ motions are more ``forced''; 
    the two tubes move largely in a synchronous fashion for the $S_x$ pattern,
    whereas the flows in between the two tubes periodically ``collide''
    when the $A_x$ pattern is examined. 
The same argument applies here although elliptic tubes are addressed, 
    with the insets of Fig.~\ref{fig_xmajor_2dfield}b 
    already hinting at a more forced behavior for $A_x$ motions.  
Also similar to \citetalias{2015A&A...582A.120S} is that, for a given $a/b$,
    the difference between the values of $P$ (or $\tau/P$) 
    for the $S_x$ and $A_x$ motions tends to decrease monotonically 
    with $d$.
This agrees with the intuitive expectation for a weaker tube interaction.

Let us pay more attention to the overall behavior for $P$ 
    to increase monotonically with $a/b$ for a given $d/b$.
Somehow subtle is that tube interactions also have some effect, because 
    increasing $a/b$ actually brings the two tubes effectively closer
    even if the distance between the tube centers is fixed.
Regardless, this tube interaction only plays a minor role,
    given that the monotonical $a/b$-dependence takes place 
    for different $d$ and for the $S_x$ and $A_x$ motions alike.
One may therefore expect the same $a/b$-dependence for isolated elliptic tubes
    (or equivalently $d/b\to\infty$), in which case it is no longer necessary to 
    distinguish between the $S_x$ and $A_x$ patterns. 
This was indeed seen in our previous IVP study 
    \citep{2020ApJ...904..116G}, where we offered some heuristic argument
    similar to the earlier one by \citet{2003A&A...409..287R}.
For simplicity, suppose that the equilibrium configuration is transversely structured 
    in a piece-wise constant manner.
The heuristic argument then relies on the observation that 
    the velocity normal to the tube edge plays a central role
    for an isolated density-enhanced tube to communicate with its surroundings. 
Evidently, a larger $a/b$ makes the tube edge more elongated in the $x$-direction. 
Their velocities primarily $x$-directed, the fluid parcels in the tube interior 
    therefore become less aware of the surroundings, 
    meaning some enhanced effective inertia and hence a longer period. 
The same argument can be invoked in this study, despite the subtlety that 
    the fluids surrounding one tube actually embed another tube.

\subsection{$S_x$ and $A_x$ Motions for the $x$-minor Orientation} 
\label{sec_sub_2ellip_xMn}    
This subsection addresses the $x$-minor orientation. 
We start with Fig.~\ref{fig_xminor_vxt} where the same set of physical quantities
    as in Fig.~\ref{fig_xmajor_vxt} is employed, 
    and the time sequences for 
    the $x$-speed sampled at the left tube center
    are presented in an identical format. 
A fitting procedure is once again performed for any $v_x$ sequence
    over the duration exactly
    enclosing the six extrema starting with the one indicated by 
    the pertinent asterisk.
\revise{We take the view} that the best-fit periods ($P$) and damping times ($\tau$)
    pertain to the ideal quasi-modes supported
    by our two-elliptic-tube configuration.

Figure~\ref{fig_xminor_2dfield} specializes to the case
    with $[a/b=2.5, d/b=5]$, following the same format as in Fig.~\ref{fig_xmajor_2dfield}
    to present the velocity fields (the blue arrows)
    and the spatial distributions of 
        the perturbation energy density ($\epsilon$, filled contours)
    at a representative instant.
Note that the tubes are quite some distance apart in this case.
Note also that Fig.~\ref{fig_xminor_2dfield} is a snapshot extracted
    from the attached animation.
Focusing on the velocity fields, one can safely conclude from the animation 
    that the notations of $S_x$ and $A_x$ make physical sense
    as proposed by \citetalias{2008ApJ...676..717L} for two-circular-tube systems.
For both perturbation patterns, 
    the insets in the animation further indicate that the attenuation of the internal flows
    is accompanied by the enhancement of perturbation energy density $\epsilon$ 
    in the nonuniform layers,
    with the localization of $\epsilon$ strongly mediated 
    by the \Alf\ resonance.
This latter point can be readily drawn from the close
    association of the darkest portion
    in either $\epsilon$ distribution with the white curve marking where 
    the quasi-mode frequency matches the local \Alf\ frequency.

Figure~\ref{fig_xminor_yProf} presents, 
    in a format identical to Fig.~\ref{fig_xmajor_yProf}, 
    the $y$-cuts of the $x$-speed through the left tube center
    for a number of arbitrarily chosen instants.
The aim is also to show the key role that phase-mixing plays for
    the gradual development of shearing motions with increasingly small
    scales in the nonuniform layers. 
We follow Fig.~\ref{fig_xmajor_yProf} to quantify this 
    with the aid of Eq.~\eqref{eq_vxMann}, once again 
    counting the instantaneous number $\Nextrm(t)$ 
    of the $v_x$ extrema. 
The pertinent instantaneous phase variations ($\phiac(t)$), 
    on the other hand, are the same as in Fig.~\ref{fig_xmajor_yProf}
    and read $[2.82, 4.23, 5.64]\pi$ for the examined instants.
However, somehow different is that
    the relevant values for $\Nextrm$ now may deviate
    from $\lfloor\phiac/\pi\rfloor$, attaining specifically $[3, 4, 5]$.
This deviation is readily attributable to the phase $\varphi_0$ in Eq.~\eqref{eq_vxMann}.
Conversely, this set of deviations can be readily verified to 
    constrain $\varphi_0$ to a rather narrow range of
    $[(q+0.19)\pi, (q+0.37)\pi]$ with $q=0, 1$.

Figure~\ref{fig_xminor_survey} moves on to survey 
    (a) the quasi-mode periods $P$ and (b) damping-time-to-period ratios $\tau/P$
    for a substantial range of combinations $[a/b, d/b]$.
This figure is identical in format to Fig.~\ref{fig_xmajor_survey}.
The values examined for the tube separation $d$
     consistently avoid tube overlapping, 
     and hence there is no need to constrain the ratio of 
     the semi-major to semi-minor axis ($a/b$)
     for a given $d/b$.
Examine Fig.~\ref{fig_xminor_survey}a first, from which one sees
     the expected monotonic behavior for the differences 
     between the $S_x$ and $A_x$ periods to decrease with $d/b$
     for a given $a/b$.
For a given separation $d/b$, one further sees that
     the values of $P$ tend to decrease monotonically with $a/b$ 
     for the $S_x$ and $A_x$ motions alike.
However, tube interactions remain possible to be partly responsible
     for this $a/b$-dependence, because the area between the two tubes
     actually broadens with $a/b$ even if $d/b$ is fixed. 
This tube-interaction effect can hardly be dismissed given that 
     the $S_x$ and $A_x$ values deviate more strongly 
     when $a/b$ increases for a given $d/b$. 
Having said that, it remains possible to partly explain
     the monotonic $a/b$-dependence of $P$ by invoking 
     the inertia arguments initially offered for isolated density-enhanced 
     elliptic tubes.
Note that the tube edges are now more elongated in the $y$-direction,
     whereas the fluids parcels in either tube interior remain to 
     oscillate primarily in the $x$-direction.
It therefore holds that an increase in $a/b$ makes the internal
     fluid parcels better aware of the external medium, meaning
     some reduced effective inertia and hence a shorter period.  

Consider now Fig.~\ref{fig_xminor_survey}b 
     where the damping-time-to-period ratios ($\tau/P$) are examined. 
Two features arise when $a/b \lesssim 1.5$, 
     one being that $A_x$ motions attenuate more efficiently than $S_x$ ones,
     the other being that the differences between the $S_x$ and $A_x$ values
     tend to decrease with tube separation. 
These features resemble what happens for two-circular-tube systems,
     and are expected because the cross-sectional shapes are not far from circular. 
However, the behavior of $\tau/P$ at larger $a/b$ is considerably
     more complicated than in Fig.~\ref{fig_xmajor_survey}b
     where the $x$-major orientation is addressed. 
By ``complicated'' we specifically refer
     to the nonmonotoic $d/b$-dependence 
     of $\tau/P$ for either perturbation pattern. 
Take the case with $a/b=2.5$.
When $d/b$ increases, 
     one sees an increase followed by some decrease in $\tau/P$ for the $S_x$ motion,
     whereas $\tau/P$ tends to decrease for the $A_x$ motion. 
This latter simplicity is only apparent, because $\tau/P$ is bound to increase 
     with $d/b$ again such that the $A_x$ values eventually join their $S_x$ counterparts
     when the tubes are sufficiently separated. 
We have verified that the somehow involved behavior of $\tau/P$ 
     is not of numerical origin (see Sect.~\ref{sec_verif}).
Furthermore, it matters little as to how to choose the segment
     of a $v_x$ sequence for fitting, provided that this segment starts sufficiently 
     later than the rapid-varying phase (see Fig.~\ref{fig_xminor_vxt}).
The most likely reason is then that the perturbations in between the two tubes 
     depend on $a/b$ and $d/b$ in an intricate fashion, and hence an intricate 
     dependence on $a/b$ and $d/b$ of the efficiency of tube interactions. 
Regardless, the reason behind the $d/b$-dependence of $\tau/P$ is also behind, say, 
     the nonmonotonic $a/b$-dependence of $\tau/P$ for $A_x$ motions for a given $d/b$
     (see the dashed curves),
     and the occasional behavior for $S_x$ motions to damp more efficiently   
     (compare the black solid and dashed curves).
    
\section{Implications for Coronal Seismology}
\label{sec_disc}
\revise{This section discusses some seismological implications
   of our study.
Only axial fundamentals are considered.
We restrict ourselves to the inference of the
    axial \Alf\ time ($\Taxial=2L/\vAe$)
    and the dimensionless nonuniform layer width ($\bar{l}$).
We further assume that neighboring elliptic tubes  
    are imaged in, say, the EUV when the line of sight lies in the tube plane
    (see Fig.~\ref{fig_Schm}a), meaning that 
    the tubes may be readily mistaken as circular ones. 
}

\revise{Let us start with Figure~\ref{fig_xmajor_survey} where our $x$-major
    computations are summarized,
    and consider only the $S_x$ motion for $[a/b=2.5, d/b=5]$.}
Now assume that the relevant oscillating tubes possess
    \revise{the ``true'' dimensionless values $[\rhoi/\rhoe=3, \bar{l}=0.4]$} 
    (note that $L/b$ is immaterial when sufficiently large).
Let $\Pobs$ and $\tauPobs$ be the measured values
    of the period and damping-time-to-period ratio, 
    and we take $\tauPobs=6.39$ to comply with Fig.~\ref{fig_xmajor_survey}.
One therefore deduces a ``true'' value of $\Pobs/1.6$ for $\Taxial$, 
    which is recalled to be dimensional.       
\revise{Now suppose that the density contrast is measured to be $\rhoi/\rhoe=3$, 
    and let us} follow the customary practice
    to infer $\Taxial$ and $\bar{l}$  
    by seeing the $S_x$ motion as a kink quasi-mode in an isolated circular tube.
It matters little regarding whether one assigns $a$ or $b$ to the tube radius.
The dimensionless quasi-mode period $P/\Taxial$ depends essentially only on $\rhoi/\rhoe$,
    always attaining $\sim 1.42$ when $\bar{l}\lesssim 0.6$ 
    (see the green horizontal line in Fig.~\ref{fig_xmajor_survey}a; 
     see also Fig.~\ref{fig_1tubekink}b).
One therefore deduces a value of $\Pobs/1.42$ for $\Taxial$, which differs from 
    the true value by $\sim 13\%$.
However, it is considerably more problematic
    when $\tauPobs$ is invoked to infer $\bar{l}$.
Employing the resistive eigenmode approach, 
    we find that a $\tauPobs=6.39$ results in a $\bar{l}\approx 0.18$, 
    which underestimates the true value by $\sim 55\%$.
One may argue that this difference is not that significant; 
    after all the customary seismological practice yields the correct qualitative 
    picture that the oscillating tube possesses a thin boundary. 
Our point, however, is that care needs to be exercised when $\tauPobs$
    is put to quantitative use, 
    with the uncertainties in the deduced dimensionless layer width $\bar{l}$
    readily exceeding $\sim 50\%$ if one neglects the combined effect of 
    tube cross-sectional shapes and tube interactions.

\revise{Now move on to Fig.~\ref{fig_xminor_survey} where our $x$-minor
    computations are summarized.}
We assess what uncertainties the customary practice may introduce
     to the axial \Alf\ time $\Taxial=2L/\vAe$ and dimensionless layer width $\bar{l}$
     by neglecting the joint effects of cross-sectional shapes and tube interactions.
The procedure is identical to what we performed for the $x$-major orientation;
     we accordingly choose the $A_x$ values for $[a/b=2.5, d/b=3]$ given that
     they deviate the most from the horizontal lines. 
Consider the period first.
One infers a true value of $\Taxial=\Pobs/1.13$
     with Fig.~\ref{fig_xminor_survey}a, 
     whereas the customary practice remains to yield a 
     $\Taxial \approx \Pobs/1.42$ or equivalently 
     some underestimation by $\sim 20\%$.
This uncertainty in $\Taxial$ is larger than for the $x$-major orientation.  
Now taking $\tauPobs=3.44$ from Fig.~\ref{fig_xminor_survey}b as measured,
     one deduces with the customary practice a dimensionless layer width of
     $\bar{l} = 0.25$ or some underestimation of the true value ($\bar{l}=0.4$)
     by $\sim 38\%$. 
Somehow smaller than for the $x$-major case, this uncertainty remains substantial
    enough to corroborate our claim that the inhomogeneity scales 
    deduced with the customary practice need to be treated with caution.

\section{Summary}
\label{sec_conc} 
With coordinated transverse displacements in neighboring
   active region (AR) loops in mind, we have addressed damped kink motions in straight equilibria where two identical parallel density-enhanced tubes with elliptic cross-sections (elliptic tubes) are embedded in an ambient corona. 
Linear, ideal, pressureless MHD was adopted throughout. 
We concentrated on axially standing motions, formulating the 
   ensuing two-dimensional (2D) initial value problem (IVP) in the plane transverse to the equilibrium magnetic field.
We identified the direction connecting the tube centers as horizontal, 
   discriminating two tube orientations where either the major  
   (dubbed ``$x$-major'') or the minor (``$x$-minor'') axis is horizontally placed.
The system evolution was initiated with external velocity drivers, 
   implemented in such a way that the internal flow fields are primarily horizontal   and are either symmetric ($S_x$) or anti-symmetric ($A_x$) with respect to the vertical axis about which our equilibrium configuration is symmetric.  
The temporal evolution of the velocity perturbations at tube centers was
   of particular interest, allowing us to identify the quasi-mode stage as where the monochromatic time sequence follows an exponentially damped envelope. 
We paid special attention to how the quasi-mode periods and damping times depend
   on the tube separation and the cross-sectional aspect ratio, thereby addressing the impact on the dispersive properties of damped kink motions from the joint effects of tube interactions and cross-sectional shapes. 

Our numerical findings are summarized as follows. 
When two-circular-tube equilibria are examined as a special case, 
   the quasi-mode periods and damping times found with our IVP approach are consistent with the modal analysis by \citet{2015A&A...582A.120S}, thereby independently justifying both the analytical T-matrix approach and the numerical results therein.
We further find that the notions of resonant absorption and phase-mixing
   are not undermined when elliptic cross-sections are allowed for. 
Specifically, the nonuniform layers around tube boundaries feature both
   the development of velocity shears with increasingly small scales and
   the energy accumulation around where the local \Alf\ frequency matches the quasi-mode frequency.   
The $A_x$ motions are found to possess shorter periods and damp more rapidly
    than the $S_x$ motions for the $x$-major orientation, in which case
    the periods and damping times for both perturbation patterns tend
    to increase with the major-to-minor-axis ratio. 
When the $x$-minor orientation is addressed, 
    the $A_x$ motions remain to oscillate more rapidly than the $S_x$ ones.
However, the periods for both patterns tend to decrease monotonically
    with the major-to-minor-axis ratio. 
Somehow subtle are the joint effects of tube interactions and cross-sectional shapes
    on the damping times, which may depend non-monotonically on the tube separation or the major-to-minor-axis ratio.
Furthermore, the $A_x$ motions may occasionally damp less efficiently. 
Neglecting these joint effects may introduce 
    some uncertainty of $\sim 20\%$ ($\sim 50\%$) to the axial \Alf\ time (the inhomogeneity lengthscale) deduced with the period (damping time).
Consequently, care may need to be exercised when the damping
    times of kink motions are put to standard seismological practice built on wave theories for isolated circular tubes.  

Some further remarks seem necessary before closing. 
Firstly, we note that our IVP approach is not restricted to two-tube systems
    but capable of handling rather generic 2D inhomogeneities. 
In particular, axially standing motions may be examined 
    for straight configurations comprising multiple tubular structures with arbitrary cross-sections.   
Collective motions in multiple neighboring resolved loops will
    then be better understood, together with the motions in those
    apparently isolated loops that actually involve a multitude of unresolved strands
    (see e.g., Sect.4.2 of the review by \citealt{2014LRSP...11....4R} for more on this multi-stranded nature of AR loops).
Secondly, that kink motions are resonantly absorbed for 
    configurations with continuous transverse structuring is
    actually much-expected \citep[e.g., the review by][]{2011SSRv..158..289G}.
However, somehow largely unnoticed is that the decayless regime
    was also observed for coordinated kink motions in adjacent AR loops \citep{2012ApJ...751L..27W}.
Accept that the transverse structuring is likely to be continuous
    in the relevant configurations. 
It then becomes necessary to address what counteracts the ensuing resonant damping,
    thereby bringing the understanding of decayless kink motions in multi-tube systems closer 
    to the level of sophistication reached for the decayless regime in isolated tubes 
    \citep[e.g.,][]{2016A&A...591L...5N,2019ApJ...870...55G,2021SoPh..296..124R}.
Thirdly, coordinated motions in multiple tubular structures
    were identified not only in AR loops but also in, say, groups of prominence threads \citep[see e.g.,][for a review]{2018LRSP...15....3A}.  
To our knowledge, so far the only theoretical investigation in
    the prominence context was the one by \citet{2009ApJ...693.1601S} with the T-matrix formalism. 
An IVP perspective is expected to be as fruitful, 
    provided that one further account for, say, the effects associated with partial ionization \citep[e.g., the review by][]{2018SSRv..214...58B}.

\begin{acknowledgements}
This research was supported by the 
    National Natural Science Foundation of China
    (41974200, 
     12273019,    
     12373055, 
     12203030,    
     and 
     42230203).    
We gratefully acknowledge ISSI-BJ for supporting the international team
    ``Magnetohydrodynamic wavetrains as a tool for probing the solar corona'', and ISSI-Bern for supporting the international team 
    ``Magnetohydrodynamic Surface Waves at Earth’s Magnetosphere and Beyond''.
\end{acknowledgements}

\bibliographystyle{aa}
\bibliography{Bib_up2date}

\begin{appendix}

\section{Subtleties with Domain Size and Grid Spacing}
\label{sec_App_num}
This \revise{appendix} examines the influences that the numerical grid setup
   may have on our time-dependent solutions,
\revise{taking two-circular-tube computations as examples.
Our $[b, \bar{l}]$ notations are used together with 
   the \citetalias{2002ApJ...577..475R} convention $[\Rtrd, \ltrd]$.
It suffices to address only the $S_x$ pattern
   for a fixed combination
   of physical parameters $[\rhoi/\rhoe=5, \ltrd/\Rtrd=0.4, d/\Rtrd=2.5, kb=\pi/30]$.
} 

We start with Fig.~\ref{fig_2tubeTst_DomainSize} by displaying the temporal
   evolution of the $x$-speed at the left tube center
   ($v_x(-d/2, 0; t)$, the solid curves) for a number of domain sizes 
   $[\xM, \yM]$ as discriminated by the different colors. 
The best-fit damping envelopes are further plotted by the dashed curves for reference.
All computations pertain to a fixed grid spacing $\Delta=0.01b$, which 
   is recalled to be different from the reference value in Table~\ref{tab_solset}.
Some differences are seen among the solid curves, 
   the most prominent one being that the monochromatic cycles 
   in the $v_x$ signal are slightly delayed when the domain size increases. 
We find that this domain size effect arises because the zero-gradient boundary condition
   is not fully transparent to impinging motions
   initiated with the pair of planar perturbations given by Eq.~\eqref{eq_initPert}.
Some weakly reflected disturbances reach and interact with the composite structure,
   resulting in some longer time for the internal flow fields to settle toward a regular pattern
   for a larger domain. 
That the domain size effect is primarily attributed to the form of the initial perturbation
   is further corroborated by an additional series of computations where 
   we multiply each exponential term in Eq.~\eqref{eq_initPert}
   by a factor $\exp(-y^2/b^2)$.
The dependence on the domain size is then found to disappear in the time-dependent solutions
   responding to this localized perturbation.
A monochromatic pattern remains clearly visible after some transitory phase, the period
   being nearly identical to what we find when Eq.~\eqref{eq_initPert} is employed.
However, the damping envelope deviates considerably from an exponential one, 
   making it difficult to examine ideal quasi-modes 
   from an IVP perspective. 
The reasons for us to stick to Eq.~\eqref{eq_initPert} are twofold. 
Firstly, what drives the internal field toward a quasi-mode behavior
   is beyond our scope.
Rather, we content ourselves with deriving the relevant periods and damping times.
Secondly, varying the domain size does not influence the best-fit period ($\Pfit$)
   and only marginally impacts the best-fit damping time ($\taufit$).
In fact, $\Pfit$ always evaluates to $\sim 112.5b/\vAe$ for the examined domains.
The value of $\taufit \vAe/b$, on the other hand, changes from
    $320.1$ for $[\xM=12b, \yM=10b]$ only modestly to $313.3$ for $[\xM=18b, \yM=15b]$.
An even weaker variation is found when the domain is
    further extended to $[\xM=24b, \yM=20b]$,
    in which case $\taufit \vAe/b$ attains $314.3$.

\begin{figure*}
\centering
\includegraphics[width=.85\textwidth]{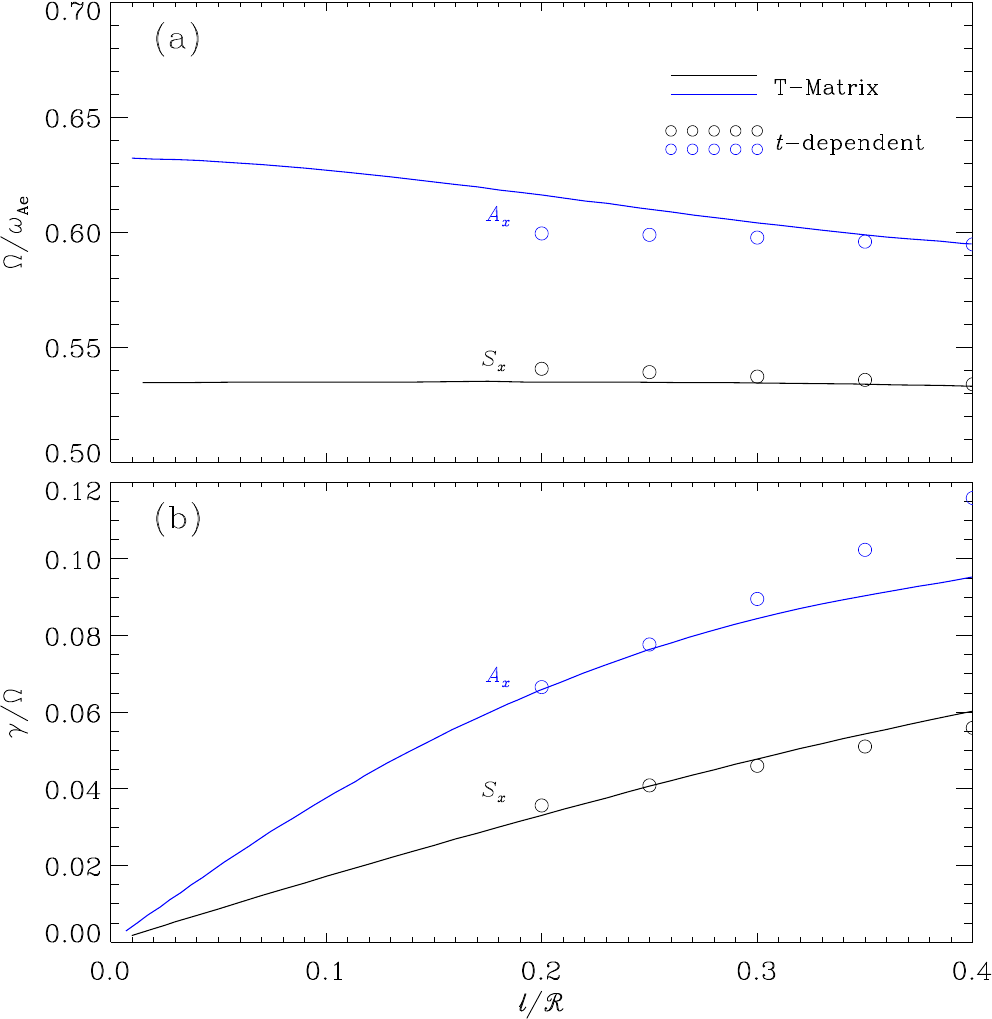}
\caption{
Comparison of (a) the oscillation frequencies $\Omega$ and 
              (b) damping rates $\gamma$ of ideal quasi-modes
    deduced from time-dependent computations (the open circles)
    with those expected with the T-matrix formalism (solid lines).
With different colors we discriminate the results for the $S_x$ and $A_x$ patterns. 
The T-matrix curves are read from Fig.~4 of 
    \citet[][\citetalias{2015A&A...582A.120S}]{2015A&A...582A.120S}.
Shown here is how $\Omega$ and $\gamma/\Omega$ depend on $\ltrd/\Rtrd$, 
    while the rest of the parameters
    is fixed at $[\rhoi/\rhoe=5, d/\Rtrd=2.5, kb=\pi/30]$. 
The $[\ltrd/\Rtrd, d/\Rtrd]$ notations follow from \citetalias{2015A&A...582A.120S}
    for the ease of comparison. 
 }
\label{fig_2tubeTst_cpTMatrix} 
\end{figure*}

\begin{figure*}
\centering
\includegraphics[width=.95\textwidth]{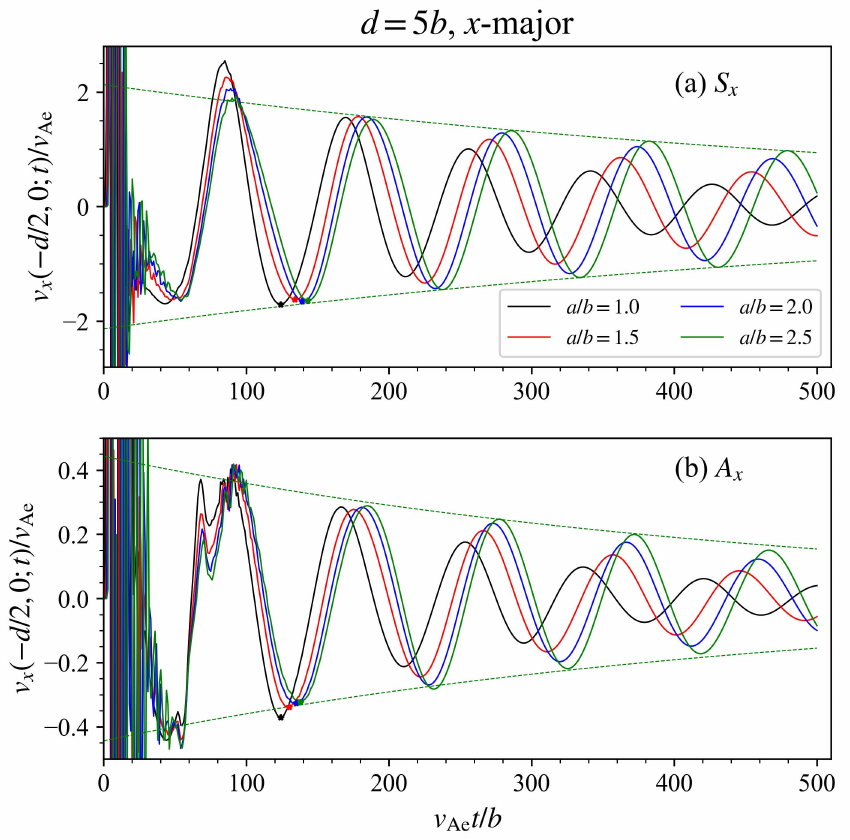}
\caption{
Kink motions with (a) $S_x$ and (b) $A_x$ patterns in 
   a two-elliptic-tube configuration with $x$-major orientation.
Plotted are the time sequences of the $x$-speed sampled at the left tube center
   ($v_x(-d/2, 0;t)$, the solid curves) for a number of ratios of
   the semi-major to semi-minor axis ($a/b$) as discriminated by the different colors. 
A fitting procedure is performed, for each sequence, over the duration
   encompassing the six extrema starting with the one denoted by an asterisk.
The best-fit damping envelope is plotted for the entire sequence, but only 
   for $a/b=2.5$ to avoid over-crowding the plots. 
The computations pertain to a fixed combination
   $[\rhoi/\rhoe=3, \bar{l}=0.4, d/b=5, kb=\pi/30]$.
}
\label{fig_xmajor_vxt} 
\end{figure*}

\begin{figure*}
\centering
\includegraphics[width=.95\textwidth]{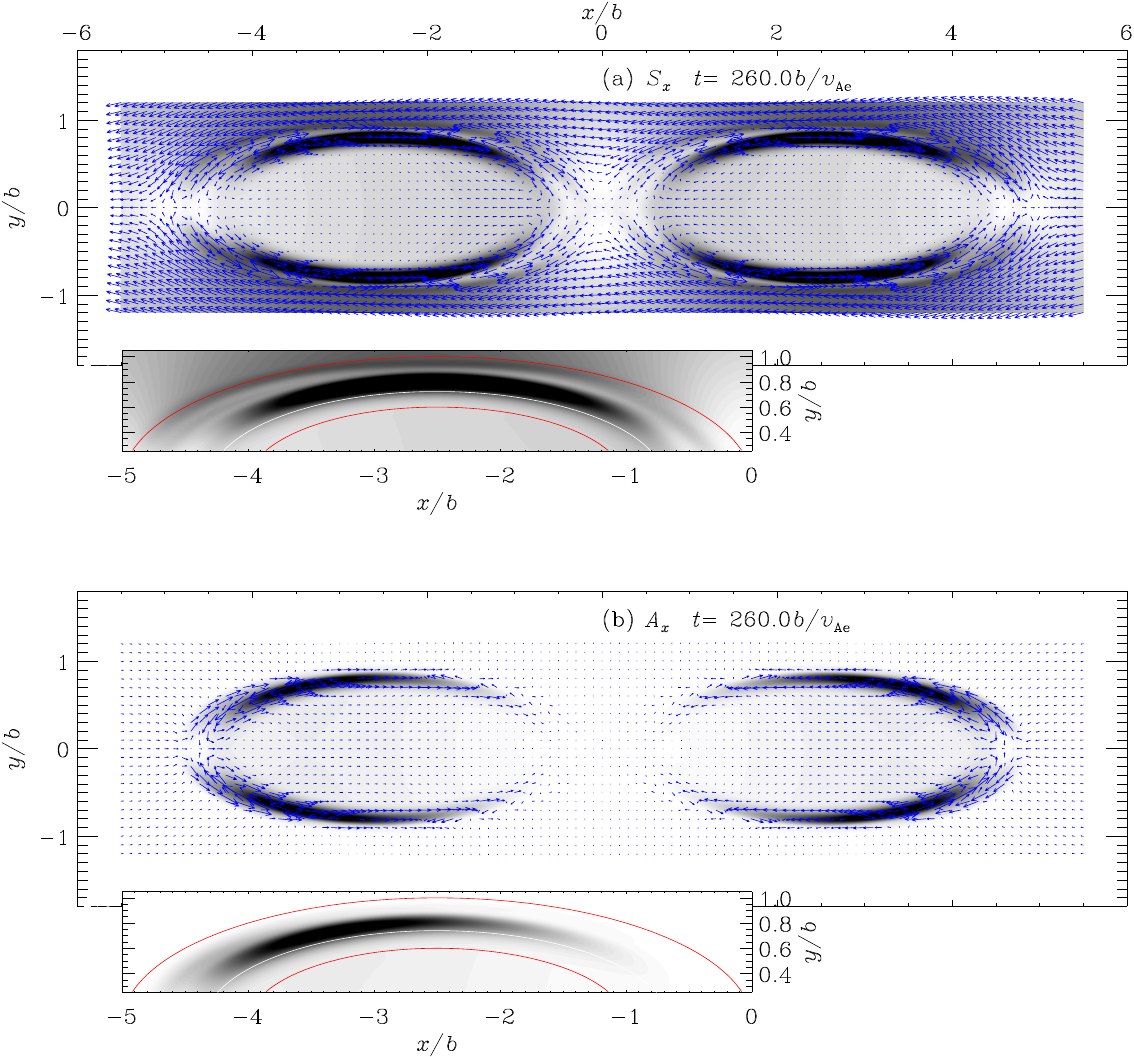}
\caption{
Representative snapshots of the velocity field (the blue arrows)
    and energy density distribution ($\epsilon$, filled contours)
    for (a) the $S_x$ motion and (b) the $A_x$ one in 
    a two-elliptic-tube configuration with the $x$-major orientation.
The inset in each panel emphasizes the distribution of $\epsilon$
    in a representative nonuniform layer, 
    whose boundaries are shown by the red solid curves.
The white contour indicates where the \Alf\ frequency $\omgA=k\vA$ 
    equals the quasi-mode frequency deduced with the fitting procedure.
Both computations pertain to a fixed combination
   $[\rhoi/\rhoe=3, a/b=2.5, \bar{l}=0.4, d/b=5, kb=\pi/30]$.
This snapshot is extracted from the animation attached to the current figure.
The $\epsilon$ contours in all snapshots are filled in such a way that
    darker portions correspond to larger $\epsilon$.
The blue arrows are also consistently scaled in all snapshots such that
    longer arrows correspond to stronger velocities.  
}
\label{fig_xmajor_2dfield} 
\end{figure*}

\begin{figure*}
\centering
\includegraphics[width=.95\textwidth]{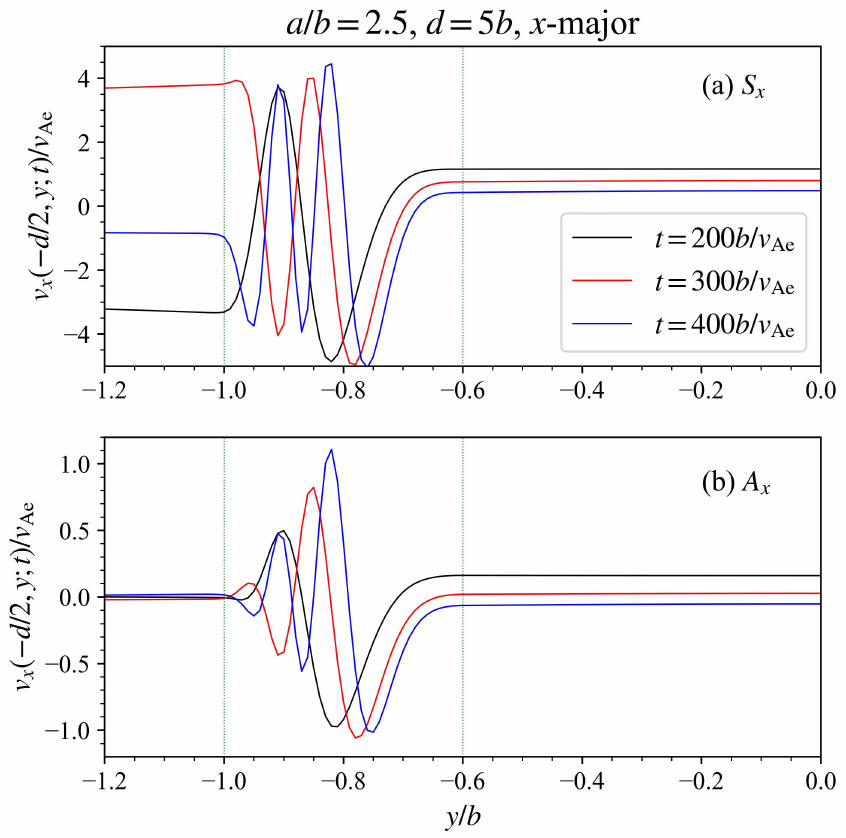}
\caption{
Distributions of the $x$-speed along the $y$-cut through the left tube center
    at a number of instants as labeled.
Panels (a) and (b) correspond to the $S_x$ and $A_x$ motions, respectively.    
The vertical dotted lines represent the borders of
    the nonuniform layer. 
Only the lower half ($y<0$) of a $y$-profile is displayed, 
    the other half being symmetric with respect to $y=0$.     
All computations are conducted for the $x$-major orientation
    with a fixed combination of physical parameters
    $[\rhoi/\rhoe=3, \bar{l}=0.4, d/b=5, kb=\pi/30]$.
}
\label{fig_xmajor_yProf} 
\end{figure*}

\begin{figure*}
\centering
\includegraphics[width=.95\textwidth]{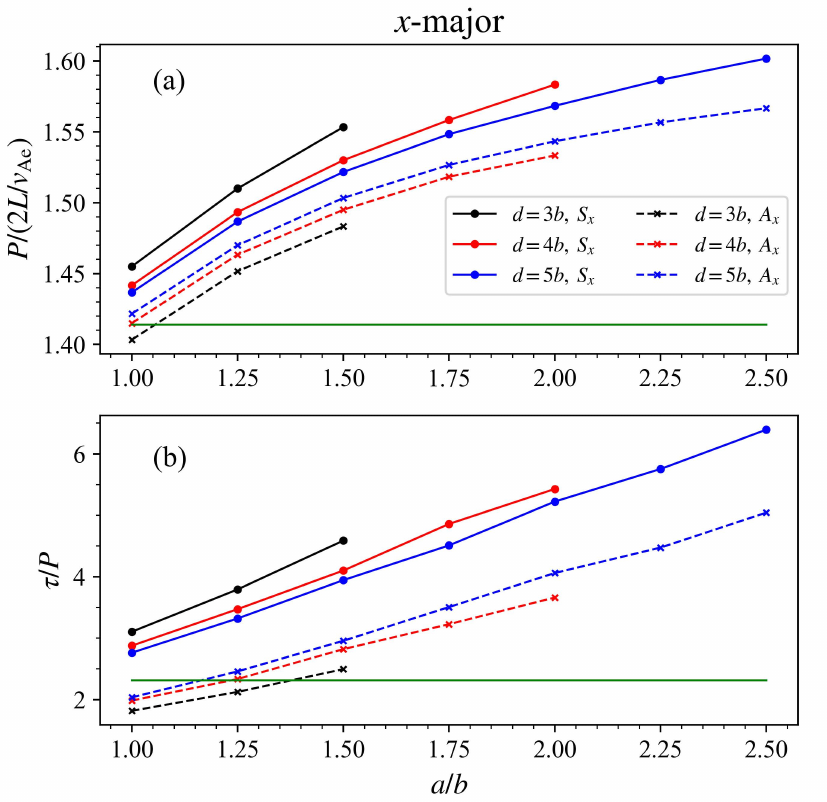}
\caption{
Dependencies of 
        (a) the quasi-mode periods ($P$) 
    and (b) damping-time-to-period ratios ($\tau/P$) 
    on the ratio of semi-major to semi-minor axis ($a/b$)
    for kink motions in a two-elliptic-tube configuration 
    with the $x$-major orientation. 
The $S_x$ and $A_x$ patterns are discriminated by the linestyles.
A number of dimensionless tube separations ($d/b$) are examined
    as differentiated by the different colors, 
    with $d/b$ constrained to avoid tube overlapping. 
All results pertain to a fixed combination of physical parameters
    $[\rhoi/\rhoe=3, \bar{l}=0.4, kb=\pi/30]$.
The horizontal lines represent the results for an isolated circular tube
    with the same set of parameters.
}
\label{fig_xmajor_survey} 
\end{figure*}

\begin{figure*}
\centering
\includegraphics[width=.95\textwidth]{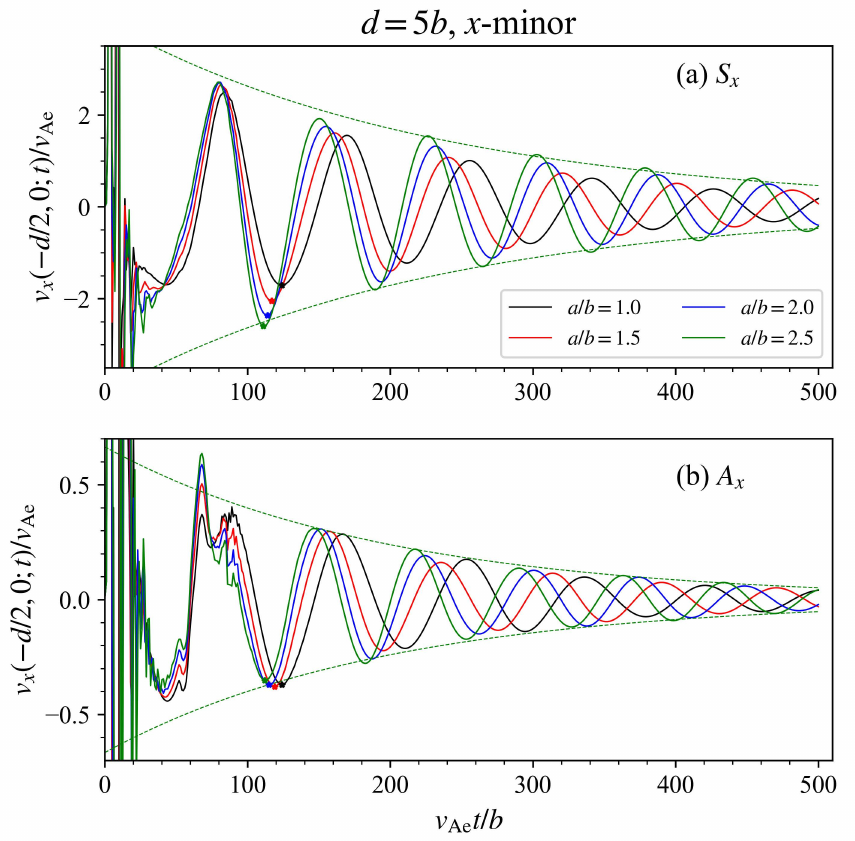}
\caption{
Similar to Fig.~\ref{fig_xmajor_vxt} but for the $x$-minor orientation.
}
\label{fig_xminor_vxt} 
\end{figure*}

\begin{figure*}
\centering
\includegraphics[width=.95\textwidth]{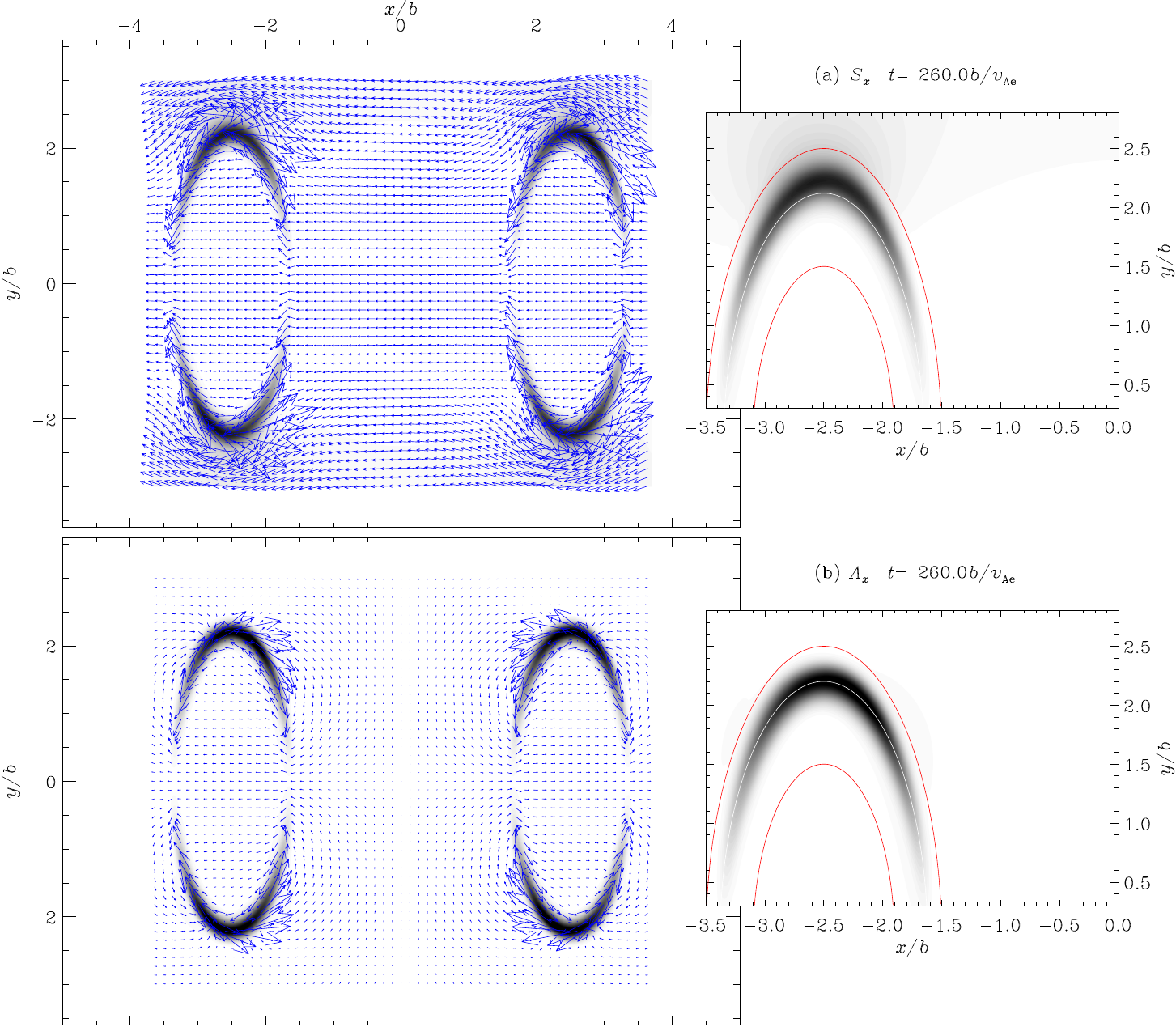}
\caption{
Similar to Fig.~\ref{fig_xmajor_2dfield} but for
    the $x$-minor orientation. 
The associated animation is also attached.      
}
\label{fig_xminor_2dfield} 
\end{figure*}

\begin{figure*}
\centering
\includegraphics[width=.95\textwidth]{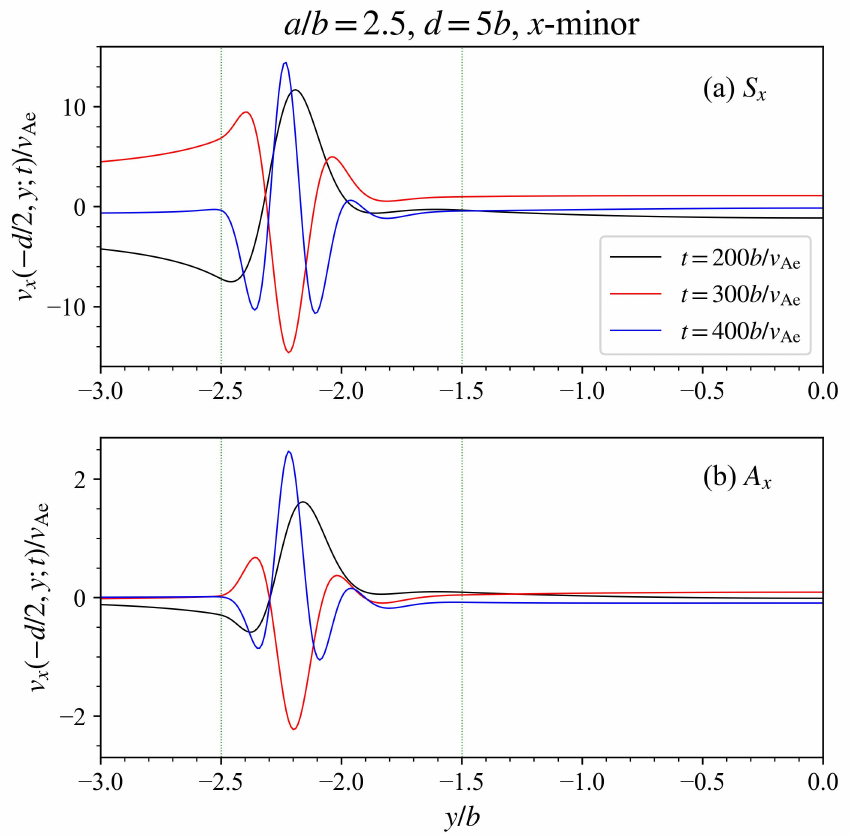}
\caption{
Similar to Fig.~\ref{fig_xmajor_yProf} but for the $x$-minor orientation.
}
\label{fig_xminor_yProf} 
\end{figure*}

\begin{figure*}
\centering
\includegraphics[width=.95\textwidth]{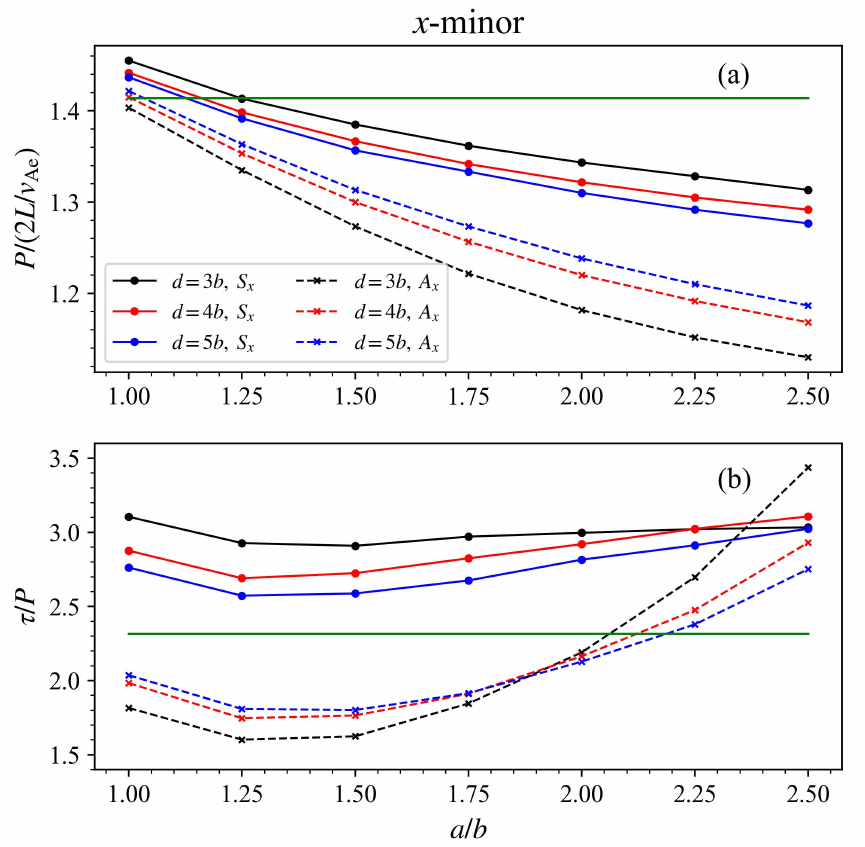}
\caption{
Similar to Fig.~\ref{fig_xmajor_survey} but for the $x$-minor orientation.
}
\label{fig_xminor_survey} 
\end{figure*}

\begin{figure*}
\centering
\includegraphics[width=.85\textwidth]{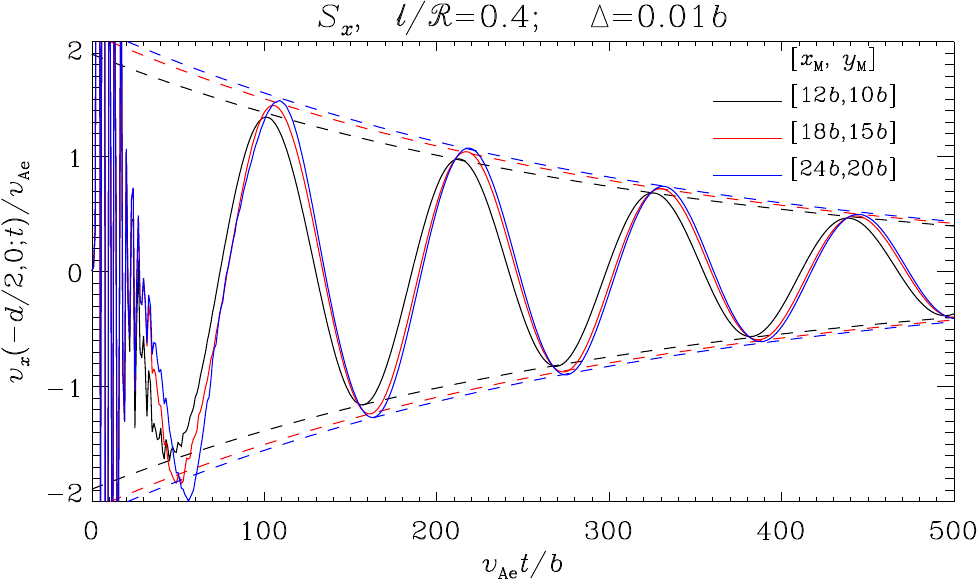}
\caption{
Temporal profiles of the $x$-speed sampled at the left tube center
   ($v_x(-d/2, 0;t)$, the solid curves)
   for a number of domain sizes as labeled. 
Overplotted by the dashed curves are the best-fit damping envelopes.
All computations pertain to the $S_x$ motion for 
    a fixed combination of physical parameters
    $[\rhoi/\rhoe=5, \ltrd/\Rtrd=0.4, d/\Rtrd=2.5, kb=\pi/30]$.
The grid spacing is fixed at $\Delta=0.01b$. 
}
\label{fig_2tubeTst_DomainSize} 
\end{figure*}

\begin{figure*}
\centering
\includegraphics[width=.85\textwidth]{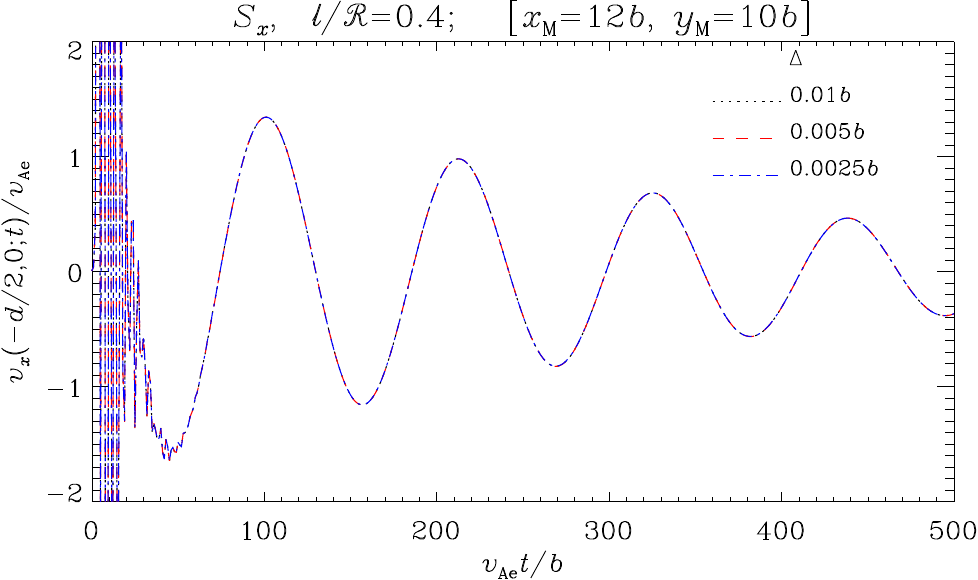}
\caption{
Temporal profiles of the $x$-speed sampled at the left tube center
   for a number of grid spacings as labeled. 
All computations pertain to the $S_x$ motion for 
    a fixed combination of physical parameters
    $[\rhoi/\rhoe=5, \ltrd/\Rtrd=0.4, d/\Rtrd=2.5, kb=\pi/30]$.
The computational domain size is fixed at $[\xM=12b, \yM=10b]$. 
}
\label{fig_2tubeTst_spacing} 
\end{figure*}

One may ask why we chose a spacing $\Delta=0.01b$
    different from the reference value ($\Delta=0.005b$)
    when examining the domain size effect.
The answer is that the specific values of $\Delta$ do not impact
    the $v_x(-d/2, 0; t)$ sequences
    for a given domain size if $\Delta$ is sufficiently small.
By ``sufficiently'' we mean some criterion related to the pair of
    equilibrium quantities $[\rhoi/\rhoe, \bar{l}]$
    (see Sect.~\ref{sec_sub_solmethod}).
We reach this statement by varying only $\Delta$
    for a substantial fraction 
    of the computations in Fig.~\ref{fig_2tubeTst_cpTMatrix}.
Consider only those with the reference domain size.     
Figure~\ref{fig_2tubeTst_spacing} offers a representative case with $\ltrd/\Rtrd=0.4$ 
    by discriminating the $v_x(-d/2, 0; t)$ sequences 
    for different $\Delta$ with the different linestyles.
Relative to the computation with the reference $\Delta$ (the red dashed curve),
    no difference can be discerned when the spacing is
    halved (blue dash-dotted) or doubled (black dotted).
The reference $\Delta$ is therefore overly demanding of computational resources
    for $\ltrd/\Rtrd=0.4$, and the same can be said even for 
    the most stringent case where $\ltrd/\Rtrd=0.2$ provided that
    only $v_x(-d/2, 0; t)$ is of interest. 
That said, varying $\Delta$ necessarily entails differences in other aspects 
    of the flow fields.
We illustrate this point by fixing $\ltrd/\Rtrd$ again at $0.4$
    and implementing the reference domain size. 
    
\begin{figure*}
\centering
\includegraphics[width=.85\textwidth]{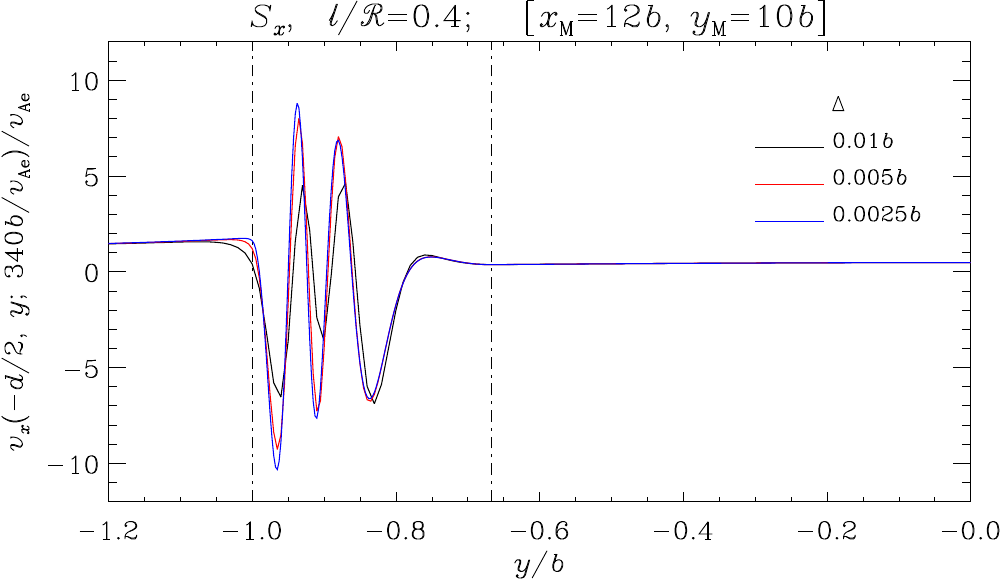}
\caption{
Distributions of the $x$-speed at $t=340b/\vAe$
    along the $y$-cut through the left tube center
    for a number of grid spacings as labeled. 
All computations pertain to the $S_x$ motion for 
    a fixed combination of physical parameters
    $[\rhoi/\rhoe=5, \ltrd/\Rtrd=0.4, d/\Rtrd=2.5, kb=\pi/30]$.
The computational domain size is fixed at $[\xM=12b, \yM=10b]$.
Only the lower half ($y\le 0$) is shown for any $y$-profile,
    the other half being symmetric about $y=0$. 
The vertical dash-dotted lines delineate the borders of
    the nonuniform layer. 
}
\label{fig_2tubeTst_yProf} 
\end{figure*}

Figure~\ref{fig_2tubeTst_yProf} presents the spatial profiles
    of the $x$-speed along the $y$-cut through the left tube center
    at $t=340 b/\vAe$.
This instant is chosen rather arbitrarily to be marginally inside
    the interval where our fitting procedure is performed. 
A cut through a tube center, on the other hand, is chosen to 
    ease the description of some key dynamics. 
A number of grid spacings are implemented and discriminated
    by the different colors. 
Only the lower half ($y\le 0$) of a $y$-profile is displayed because 
    the other half is symmetric with respect to $y=0$. 
The vertical dash-dotted lines represent the borders
    of the nonuniform layer.
Focus for now on this layer, where one sees a series of ripples in all computations.
The most prominent effect associated with $\Delta$ is that
    the ripples are better resolved when the numerical grid gets finer,
    with the computed magnitudes of the local extrema being larger.
Physically speaking, however, all computations prove capable of capturing 
    the phase-mixing physics pertaining to the \Alf\ continuum. 
To demonstrate this, we note that the \Alf\ frequency $\omgA=k \vA$
    is locally symmetric about $x=-d/2$, meaning $\partial\omgA/\partial x=0$
    for the chosen cut. 
It then follows from \citet{1995JGR...10019441M} that the $v_x$ profile 
    in the layer can be written as
\begin{equation}
    v_x \propto \mathcal{A}(y) \cos[\omgA(y) t+\varphi_0],
    \label{eq_vxMann}
\end{equation}
    where $\mathcal{A}(y)$ is some slow-varying envelope. 
Let $\omgAi=k\vAi$ be the internal \Alf\ frequency.    
Some instantaneous phase variation $\phiac(t)=(\omgAe-\omgAi)t$ 
    is therefore accumulated over the continuum, leading to 
    $N_{\rm hc} = \lfloor\phiac(t)/\pi\rfloor$ half-cycles
    with $\lfloor\cdot\rfloor$ 
    \revise{being the floor function
    \citep[see][for similar applications]{2024arXiv240211181S}}.
Consequently, one expects $N_{\rm hc}$ or $N_{\rm hc}+1$ extrema, 
    where the uncertainty derives from the phase $\varphi_0$ in Eq.~\eqref{eq_vxMann}.
Plugging in the numbers yields a $\phiac/\pi$ of $6.27$, 
    and therefore $6$ or $7$ extrema are expected.
This expectation is faithfully reproduced in Fig.~\ref{fig_2tubeTst_yProf}.
The same can actually be said for all instants when the internal field
    behaves like a quasi-mode.
Now move on to the portions outside the nonuniform layer.
The only difference that one can barely tell for different $\Delta$
    appears in the external flow fields immediately adjacent to the layer. 
On the other hand, no difference can be discerned in the internal motions. 

\begin{figure*}
\centering
\includegraphics[width=.85\textwidth]{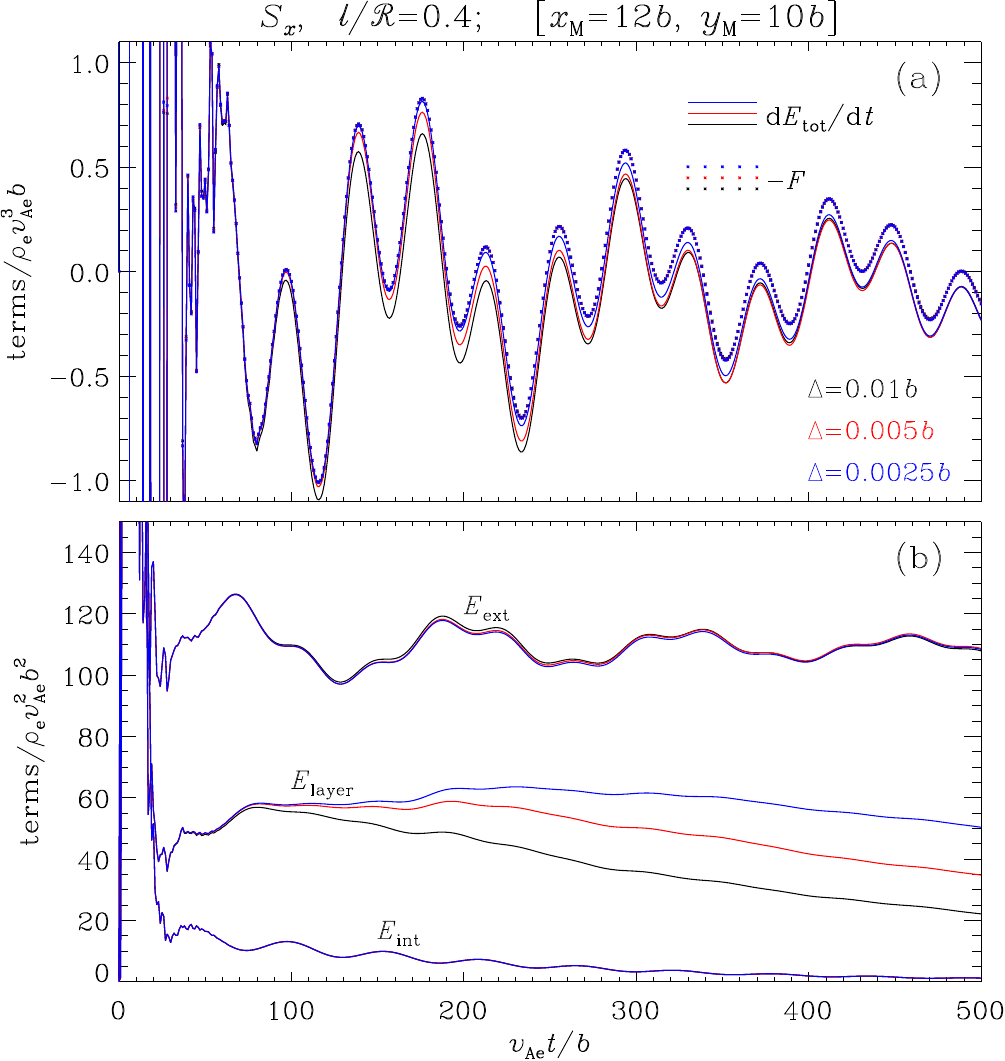}
\caption{
Temporal profiles of some energetics-related quantities
    for several grid spacings as differentiated
    by different colors. 
Energetics are consistently examined for a box 
    $[-(d/2+1.2b),(d/2+1.2b)]\times[-1.2b,1.2b]$, in which 
    the three mutually exclusive constituents are the tube interiors,
    nonuniform layers, and exterior. 
Shown in (a) are the time derivative of the total perturbation
    energy in the box ($\mathd E_{\rm tot}/\mathd t$, the sold curves),
    together with the instantaneous energy flux into 
    this box ($-F$, asterisks).
Further plotted in (b) are the instantaneous total energy 
    in the interiors ($E_{\rm int}$), 
    nonuniform layers ($E_{\rm layer}$),
    and the exterior ($E_{\rm ext}$). 
All computations pertain to the $S_x$ pattern for 
    a fixed combination of physical parameters
    $[\rhoi/\rhoe=5, \ltrd/\Rtrd=0.4, d/\Rtrd=2.5, kb=\pi/30]$.
The computational domain size is fixed at $[\xM=12b, \yM=10b]$.
}
\label{fig_2tubeTst_ener} 
\end{figure*}

Figure~\ref{fig_2tubeTst_ener} examines the grid spacing effect from 
    the energetics perspective, differentiating $\Delta$ by the different colors.
Equations~\eqref{eq_enercons} to \eqref{eq_defeps} are employed to 
    evaluate all energetics-related quantities, 
    and we take the area $Q$ therein to be a box
    $[-(d/2+1.2b), (d/2+1.2b)]\times [-1.2b, 1.2b]$. 
Figure~\ref{fig_2tubeTst_ener}a performs a gross energy balance analysis
    over $Q$ by presenting the temporal profiles of the time-derivative 
    of the instantaneous total energy ($\mathd E_{\rm tot}/\mathd t$, the solid curves)
    and those of the instantaneous energy flux into $Q$
    ($-F$, asterisks). 
The asterisks of different colors cannot be told apart, the reason being that
    all computations yield an identical set of dependents 
    at the boundaries of $Q$. 
Some difference nonetheless shows up among the solid curves,
    despite that $\mathd E_{\rm tot}/\mathd t$ should balance $F$
    (see Eq.~\ref{eq_enercons}).
As intuitively expected, gross energy conservation is maintained for a longer duration
    when a finer grid is employed, with some $\sim 30\%$ difference between
    $\mathd E_{\rm tot}/\mathd t$ and $-F$ staring to appear
    at $t\sim 90b/\vAe$ ($\sim 270b/\vAe$) when $\Delta = 0.01b$ ($0.0025b$).
Note that the examined $Q$ is the union of three mutually exclusive portions,
    namely the tube interiors (to be denoted as ``int''),
    nonuniform layers (``layer''), and exterior (``ext'').
Figure~\ref{fig_2tubeTst_ener}b then displays, against time,
    the energies in these individual portions.
By far the most significant effect that the grid spacing has
    on the energetics is that the energy in the nonuniform layer is increasingly
    under-estimated when $\Delta$ increases. 
\revise{However, there exists} only some weak difference among the $E_{\rm ext}$ 
    curves because only the flows in the immediate neighborhood 
    of either tube are somehow impacted by choices of $\Delta$. 
Furthermore, the internal energies ($E_{\rm int}$) consistently remain the same
    when $\Delta$ is varied, reinforcing the notion that $\Delta$ has no influence
    on the internal flow field. 
\end{appendix}

\clearpage

\clearpage

\clearpage

\clearpage

\clearpage

\clearpage

\clearpage

\clearpage

\clearpage

\clearpage

\clearpage

\clearpage

\clearpage

\clearpage

\clearpage

\clearpage

\end{document}